\documentclass[preprint]{aastex}      % Comments after  % are ignored
\usepackage{appendix}
\usepackage{amsmath,amssymb,amsfonts} % Typical maths resource packages
\usepackage{graphics}                 % Packages to allow inclusion of graphics
\usepackage{color}                    % For creating coloured text and background
\usepackage{hyperref}                 % For creating hyperlinks in cross references
\usepackage{natbib}
\usepackage{epsfig}
\usepackage{graphicx}                    %   for imported graphics
\usepackage{epstopdf}

\shorttitle{Secondary EUV Waves}
\shortauthors{Olmedo et al.}

\begin{document}
\title{Secondary Waves, and/or the ``Reflection'' From and ``Transmission'' Through a Coronal Hole of an EUV Wave Associated With the 2011 February 15 X2.2 Flare Observed With SDO/AIA and STEREO/EUVI}

\author{Oscar Olmedo \altaffilmark{1,5}, Angelos Vourlidas \altaffilmark{1}, Jie Zhang \altaffilmark{2}, Xin Cheng \altaffilmark{3,4}}

\affil{$^1$ Space Science Division, U.S. Naval Research Laboratory, Washington, DC 20375, USA}\email{oscar.olmedo.ctr@nrl.navy.mil}
\affil{$^2$ School of Physics, Astronomy and Computational Sciences, George Mason University, 4400 University Drive, Fairfax, VA 22030, USA}
\affil{$^3$ School of Astronomy and Space Science, Nanjing University, Nanjing 210093, China}
\affil{$^4$ Key Laboratory for Modern Astronomy and Astrophysics (Nanjing University), Ministry of Education, Nanjing 210093, China}
\altaffiltext{5}{NRC Research Associate at NRL}

\begin{abstract}
For the first time, the kinematic evolution of a coronal wave over the entire solar surface is studied. Full Sun maps can be made by combining images from the \textit{Solar Terrestrial Relations Observatory} satellites, \textit{Ahead} and \textit{Behind}, and the \textit{Solar Dynamics Observatory}, thanks to the wide angular separation between them. We study the propagation of a coronal wave, also known as ``EIT'' wave, and its interaction with a coronal hole resulting in secondary waves and/or reflection and transmission. We explore the possibility of the wave obeying the law of reflection of waves. In a detailed example we find that a loop arcade at the coronal hole boundary cascades and oscillates as a result of the EUV wave passage and triggers a wave directed eastwards that appears to have reflected. We find that the speed of this wave decelerates to an asymptotic value, which is less than half of the primary EUV wave speed. Thanks to the full Sun coverage we are able to determine that part of the primary wave is transmitted through the coronal hole. This is the first observation of its kind. The kinematic measurements of the reflected and transmitted wave tracks are consistent with a fast-mode MHD wave interpretation. Eventually, all wave tracks decelerate and disappear at a distance. A possible scenario of the whole process is that the wave is initially driven by the expanding coronal mass ejection and subsequently decouples from the driver and then propagates at the local fast-mode speed.
\end{abstract}


\maketitle
\section{INTRODUCTION}
Coronal waves are large scale bright fronts that propagate over the solar surface, with speeds ranging from 50 to over 700 km s$^{-1}$ \citep[e.g.][]{Thompson_09,Warmuth_11}. These features have traditionally been called ``EIT waves'', because they were first studied in detail \citep{Moses_97,Thompson_98,Thompson_99} with the Extreme Ultraviolet Imaging Telescope \citep[EIT;][]{Delaboudiniere_95}. It is more appropriate to call them ``EUV waves'' because they are observed in coronal extreme ultra-violet (EUV) wavelengths. ``EUV waves'' have been associated with similar phenomena in other wavelengths, such as chromospheric He I \citep{Vrsnak_02,Gilbert_04a,Gilbert_04b}, X-rays \citep{Khan_02,Narukage_02,Hudson_03,Warmuth_05,Vrsnak_06}, and radio wavelengths \citep{Klassen_00,Pohjolainen_01,Khan_02,Warmuth_04,Vrsnak_05,White_05,Vrsnak_06}. They are also associated with coronal mass ejections (CMEs), flares, and type II radio burst  \citep{Biesecker_02,Cliver_05}. Recent reviews on this topic can be found in e.g. \citet{Wills-Davey_09} \citet{Gallagher_10}, \citet{Warmuth_10}, \citet{Zhukov_11}, and \citet{Patsourakos_12}. For the remainder of this paper we will refer to this phenomena as a ``coronal wave'' or ``wave front'', which is appropriate because of their wave nature.

The interpretation of these waves has been a highly debated topic. Two broad scenarios have been proposed, either as a true wave (MHD wave propagating in the corona) or a pseudo-wave (a propagating front of an expanding structure, such as a CMEs outer envelope). Moreton waves \citep{Moreton_60} observed in H$\alpha$ have been interpreted as the chromospheric counterparts of fast-mode magneto-hydro-dynamic (MHD) coronal waves \citep{Uchida_68}. Hence, it was natural to associate Moreton waves and ``EIT waves'' \citep[e.g.][]{Warmuth_04}. Much evidence has been put forth for the MHD wave interpretation, as a fast-mode wave  \citep[e.g.][]{Wills-Davey_99,Wang_00,Wu_01,Ofman_02,Patsourakos_09a,Patsourakos_09b,Schmidt_10}, a slow-mode wave \citep{Wang_09}, or soliton wave \citep{Wills-Davey_07}. The pseudo-wave interpretation proposes several mechanisms all associated with the large-scale structure of the associated CME. One such pseudo-wave is a propagating reconnection front \citep{Attrill_07a,Attrill_07b,vanDrielGesztelyi_08}, in which as the CME is expanding it interacts with the smaller scale loops in the vicinity via reconnection to produce heat and enhanced EUV intensity. Another possible pseudo-wave mechanism is the disk projection of large-scale current shells that surround the expanding flux-rope core of a CME \citep{Delannee_08}. Joule heating and/or plasma compression can occur at these shells and lead to enhanced intensities. Yet another pseudo-wave mechanism is the plasma compression due to the successive openings of the overlying magnetic fields as the CME is expanding \citep{Chen_02,Chen_05}. \citet{Patsourakos_09b} compiled a list of several observational tests for comparing the wave and pseudo-wave explanations, upon applying these test to one event they concluded in favor of a fast-mode wave interpretation and state that this wave is ``probably'' triggered by the expansion of the loops associated with the CME, in other words that it is driven by the CME. Finally, another point of view takes on a hybrid of wave and pseudo-wave interpretations \citep{Chen_02,Chen_05,Zhukov_04,Cohen_09,Chen_11,Downs_11,Cheng_12}. In this hybrid view a CMEs outer envelope is the pseudo-wave that can drive a fast-mode wave ahead, which subsequently evolves freely over the solar surface once the CME has propagated sufficiently away from the Sun. This emerging hybrid view shows the most promise for a complete description of coronal waves as it can account for most (if not all) observational features.

The controversy seems to stem largely from ambiguities in the observed features due to limited spatial or temporal sampling. Recent ultra high cadence observations show clearly the formation of a shock in the EUV corona \citep{Liu_10, Kozarev_11,Ma_11,Cheng_12} providing strong support for the fast-mode interpretation. Further, the reflection of coronal waves at coronal hole (CH) boundaries is especially difficult to reconcile with a pseudo-wave mechanism. The first reported observation of a reflection of a coronal wave was associated with the event of 2007 May 19 \citep{Long_08,Veronig_08,Gopalswamy_09}, and more recently the 2011 June 7 event \citep{LiT_12}. \citet{Gopalswamy_09} studied the 2007 May 19 coronal wave event in detail and reported the reflection of the primary wave in three directions from a nearby CH  west of the flaring active region, and possibly a fourth by a coronal hole to the south. To perform this analysis they used running differenced images to show this effect (the same technique was also used by \citet{LiT_12} to show an example of reflection). In the technique of running difference each image in a sequence of images is subtracted from the previous image to enhance moving features or intensity variations. The use of this technique caused some controversy because it can produce misleading wave-like effects. Upon reanalysis of this event, \citet{Attrill_10} showed that using running difference caused such misleading effects, or ``optical illusions'', and further comparison with a base differencing technique concluded that the ``wave'' does not in fact reflect from the coronal hole boundary. In base difference one selects a pre-event image and systematically subtracts that image from all the images in the sequence. \citet{Zhukov_11} comments that using this technique is also flawed in the case of a wave reflected back towards the eruption site because the background intensity over which the reflected wave propagates may have changed due to EUV dimming making the wave more difficult to detect. To explain what \citet{Gopalswamy_09} reports as ``reflection'' \citet{Attrill_10}  invokes hot plasma channelled along coronal loops in the opposite direction of the ``wave'', a secondary event (presumably occurring very soon after the initial event, as suggested by the associated double-CME event), and/or a rotation of the ``wave'' due to local inhomogeneities of the fast-mode MHD wave speed. Therefore, the question of reflection remains open, especially since the techniques used to identify reflection have been called to question. Here we provide new observations of an event in which reflection could be a possible explanation, as the trajectory of the reflected wave seems to obey the law of reflection. In one example we show that there is a spatial and temporal correlation between the wave, which appears to have reflected from the CH boundary, and the oscillation of a loop arcade at the CH boundary aligned with the trajectory of the reflected wave. The loops in the arcade are triggered to oscillate one by one as the primary (or incident) wave reaches each individual loop thereby giving an appearance that the oscillation of the first loop triggers the oscillation of the next loop and so forth in what appears to resemble a cascade. However, the observations cannot conclusively determine whether the wave is in fact a true reflection or launched by the loop arcade. Excitation of coronal loop oscillations by the primary coronal wave has been reported in the literature \citep[e.g.][]{Wills-Davey_99,Ofman_02,Hudson_04,Ofman_07,Aschwanden_11}, though what is novel about this observation is the fact that an entire loop arcade oscillates and that the cascading speed was found to be related to the oblique angle at which the primary wave reached the arcade.

In addition, we report the first detection of wave transmission through a CH, as previous observations seemed to indicate that coronal waves were stopped at the boundaries of CHs\citep[e.g.][]{Thompson_98,Veronig_06}. Both reflection and transmission have been predicted in recent MHD simulations of fast-mode coronal waves \citep{Schmidt_10}. \citet{Schmidt_10} identify in their simulations ``secondary waves'', which seem to describe both reflection and transmission waves. For the sake of generality, we use the same approach and define a ``secondary wave'' as any other observed wave induced (either directly or indirectly) by the primary wave including apparent reflected and transmitted waves. For example, we call the wave triggered by the cascading loop arcade at the CH boundary a secondary wave.

Our paper is organized as follows. In \S~\ref{sec_obs}, we introduce the multi-view point and multi-wavelength space based EUV observations of the Sun and the full Sun EUV maps we use for analyzing the wave propagation on a global scale. In \S~\ref{sec_results}, we discuss our kinematic analysis of the wave reflection and transmission. We conclude in \S~\ref{sec_conclusion}.

\section{OBSERVATIONS AND METHODOLOGY}\label{sec_obs}

In this paper we study the coronal wave associated with the X2.2 flare and CME of 2011 February 15, with NOAA-reported$^[$\footnote{NOAA 2011 events: \url{ftp://ftp.ngdc.noaa.gov/STP/SOLAR_DATA/SOLAR_FLARES/FLARES_XRAY/2011/XRAY2011}}$^]$ flare onset time of 01:44 UT and peak time of 01:56 UT.  This event has also been studied by other authors who have looked into the photospheric, magnetic, and helioseismic evolution during the flare \citep{Kosovichev_11,Schrijver_11,Beauregard_12,Gosain_12,Jiang_12,Petrie_12,Wang_12}. An extensive study on this event was recently made by \citet{Schrijver_11} who discussed the coronal wave thermal properties, coronal loop evolution, and CME eruption and its connection to the coronal wave. The primary focus of the work of \citet{Schrijver_11} was the initial stages of the eruption (up to $\sim$02:00 UT), and based on a zero-$\beta$ MHD simulation studied the pseudo-wave component of this event, namely the outer envelope of the expanding CME, which they interpret to be a current-shell \citep{Delannee_08}. Taking the hybrid interpretation, the work presented herein will focus on the freely propagating fast-mode wave component initially driven by the CME. This work, together with the work of \citet{Schrijver_11}, give a complete hybrid picture of the coronal wave associated with this event.

\begin{figure}
    \centerline{\hspace*{0.0\textwidth}
   \includegraphics[width=0.6\textwidth]{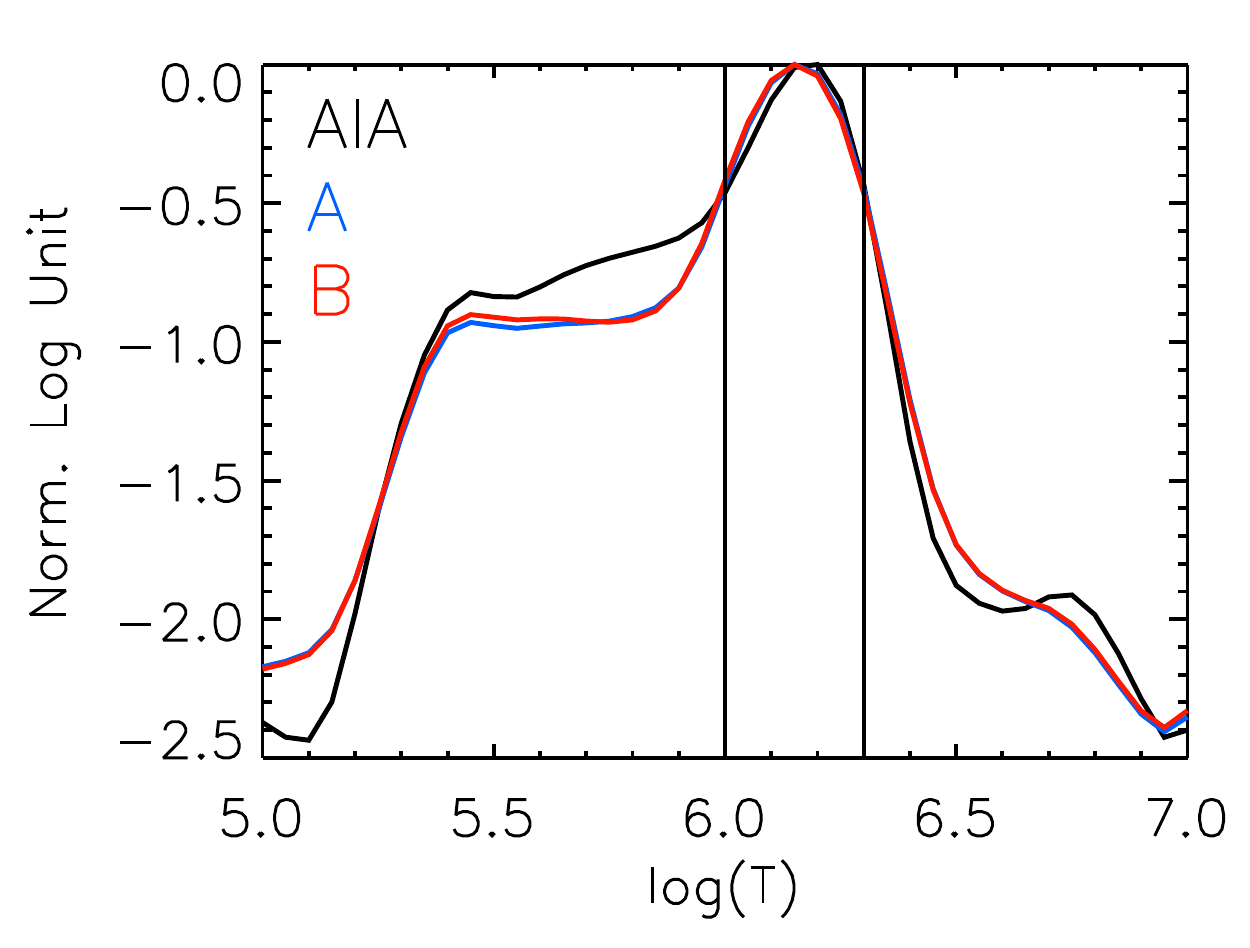}
   }
   \caption[AIA \& EUVI temperature response curves]{The temperature response of the AIA 193 {\AA}, EUVI-A 195 {\AA}, and EUVI-B 195 {\AA} passbands. In peak-normalized log units. The vertical lines at 6.0 and 6.3 correspond to 1 and 2 MK respectively.}
   \label{fig_tres}
\end{figure}

We combine observations from the Extreme Ultraviolet Imaging Telescope \citep[EUVI;][]{Wuelser_04} aboard the \textit{Solar Terrestrial Relations Observatory} \citep[STEREO;][]{Kaiser_08}, part of the Sun-Earth Connection Coronal and Heliospheric Investigation \citep[SECCHI;][]{Howard_08}, and the Atmospheric Imaging Assembly \citep[AIA;][]{Lemen_11} aboard the \textit{Solar Dynamics Observatory} (SDO). These three EUV full disk imagers give us the capability to image the entire EUV Sun instantaneously. We make maps of the full Sun by combining AIA 193 {\AA} images, taken from the vantage point at Earth, with EUVI 195 {\AA} observations from the vantage points of the twin STEREO satellites AHEAD (A), and BEHIND (B). Figure \ref{fig_tres} shows the temperature response of the three instruments. The black line corresponds to AIA 193 {\AA}, the blue to EUVI-A 195 {\AA}, and the red to EUVI-B 195 {\AA}. The response of EUVI-A and EUVI-B are essentially identical. Comparing AIA with EUVI, there is good overlap over 1-2 MK, with peak temperatures of 1.6 MK and 1.5 MK, respectively. It is known that coronal waves can be clearly observed in a temperature range between 1 and 2 MK \citep[e.g.][]{Patsourakos_09b}, therefore these combined observations are well suited for studying coronal waves.

The STEREO A satellite orbits ahead of the Earth and the STEREO B behind the Earth, both with orbit radius approximately equal to 1 AU around the Sun. Immediately before the time of the event, at 01:35 UT on 2011 February 15, the separation angle between the Earth and the STEREO A satellite was $86.75^\circ$ and the STEREO B satellite was $94.08^\circ$. The separation angle between the STEREO satellites was therefore $\sim180^\circ$. With the STEREO observations alone it is possible to make maps of the entire Sun, with the drawback that the limb observations will suffer from projection effect, due to the curvature of the Sun. Therefore, on full Sun maps where the limbs of STEREO A and B overlap (or the ``seams''), the quality of the observations of low-lying features will be relatively poor. We can overcome this problem, at least for the ``seam'' aligned with the Earth, by replacing it with observations from AIA. A map of the Sun using these three observations is made by interpolating the images, which are in helioprojective coordinates, to heliographic coordinates, assuming that the radius is equal to one solar radius, and then onto a longitude/latitude Cartesian grid \citep{Thompson_06}. In regions on the map where multiple viewpoints overlap, the observation at each latitude/longitude position that is closes to the Sun center within each respective instruments FOV, in helioprojective coordinates, is chosen. The seams are the exact position where the distance to the Sun center within the FOV of the two instruments, in helioprojective coordinates, are equal.

\begin{figure}
   \centerline{\hspace*{0.0\textwidth}
   \includegraphics[width=0.33\textwidth]{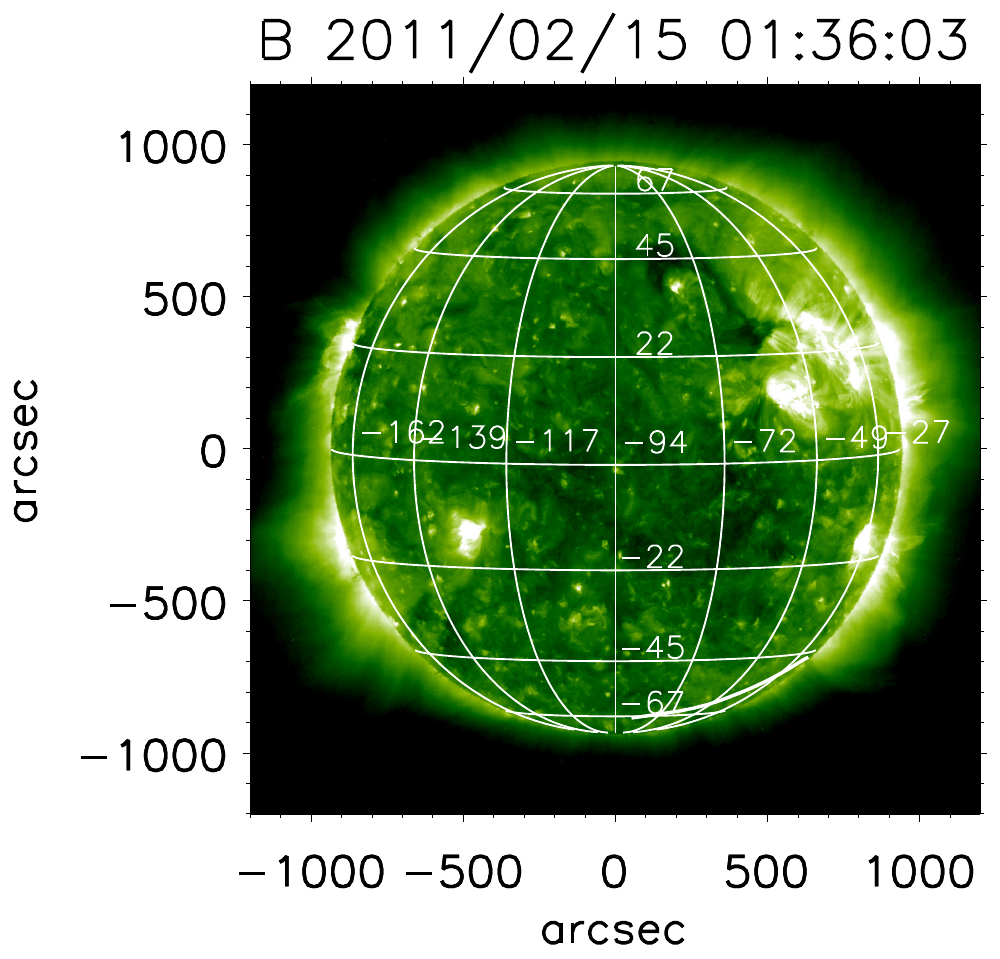}
              \includegraphics[width=0.33\textwidth]{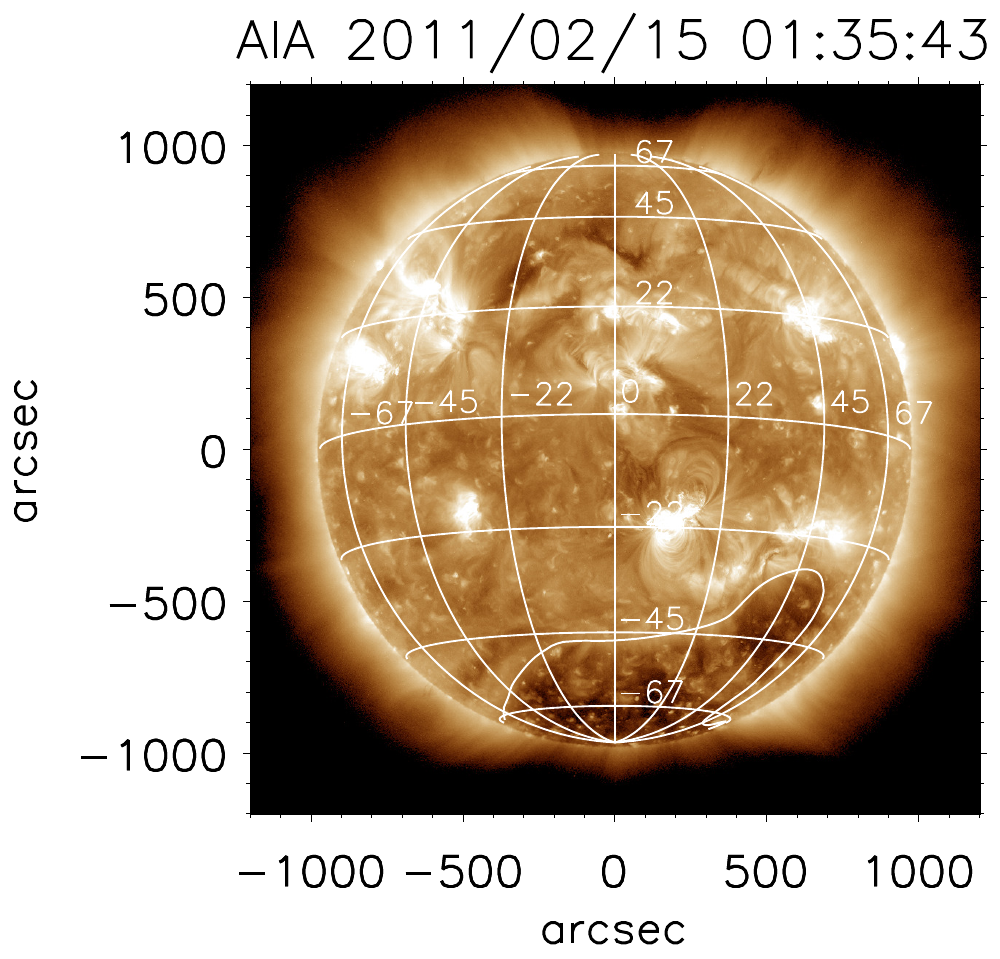}
                            \includegraphics[width=0.33\textwidth]{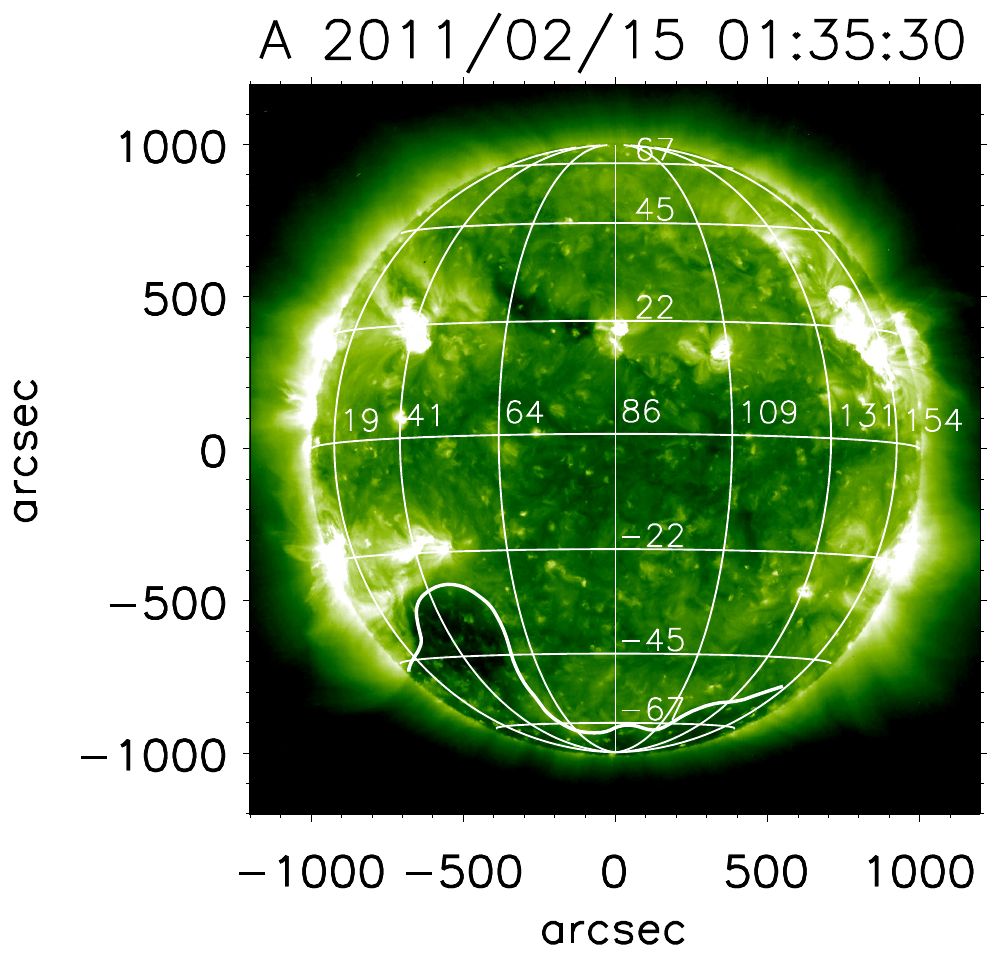}
   }
    \centerline{\hspace*{0.0\textwidth}
   \includegraphics[width=1\textwidth]{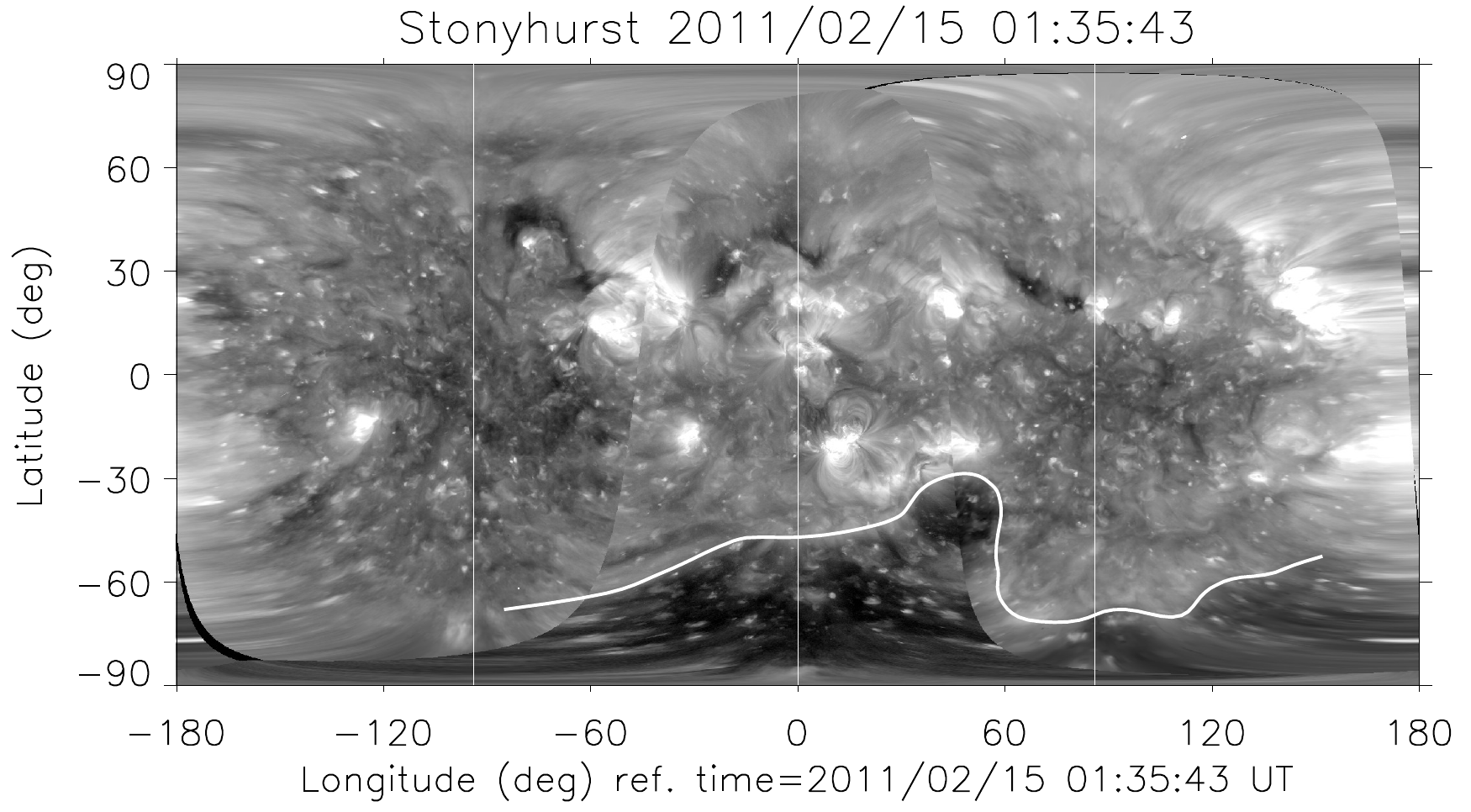}
   }
   \caption[Full Sun Map]{Combining the three observations from AIA 193 {\AA} (top center), EUVI-A 195 {\AA} (top right), and EUVI-B 195 {\AA} (top left) to make a full Sun map (bottom). Top panel shows the Sun from the three viewpoints at approximately the same reference time of 2011 February 15 $\sim$01:35 UT. The bottom panel is the map of the combined three images in Stonyhurst units. The coronal hole boundary, as outlined using the AIA 193 {\AA} image, is projected onto the map. The vertical white lines in the map correspond to the central meridians of EUVI-B, AIA, and EUVI-A with Stonyhurst Longitude at 01:35 UT of $-94^\circ$, $0^\circ$, and $86^\circ$ respectively.}
   \label{fig_fullsun}
\end{figure}

The full Sun maps are then transformed into Stonyhurst coordinates with a fixed reference time. This is the same as shifting Carrington coordinates such that the longitude of the solar central meridian as seen from the Earth, at the fixed reference time, is centered on the map and corresponds to zero degrees. Note that Stonyhurst and Carrington latitudes are equal. The longitudinal axis is in Stonyhurst units from $-180^\circ$ to $+180^\circ$. All subsequent frames are de-rotated to be aligned with the reference time. In this way, long-lived features (such as active regions, and CH's) with respect to the event will remain fixed on the map as the Sun rotates. For the 2011 February 15 event the reference time is taken to be 01:35 UT and the solar central meridian Carrington longitude ($L0$) is $+21.9^\circ$. Figure \ref{fig_fullsun} shows the transformation from the individual AIA, EUVI-A, and EUVI-B images (top row) to the map (bottom). The flare kernel site's location is at Stonyhurst longitude $12.1^\circ$ (Carrington longitude $34^\circ$) and latitude $-20^\circ$. Overlaid onto the AIA, EUVI-A, and EUVI-B FOV is a longitude/latitude grid. The vertical lines on the map (bottom) correspond to the central meridians, at 01:35 UT, of AIA ($0^\circ$), EUVI-A ($+86^\circ$), and EUVI-B ($-94^\circ$). For the 2011 February 15 event a new map is made every 5 minutes starting at 01:35 UT until the end of the event, which we define as the time when the propagating wave becomes to faint to track ($\sim$03:35UT). When making these maps we are limited to the lowest cadence of the three instruments. EUVI typically has a cadence between 2.5 and 10 minutes for normal operations, and AIA a cadence of 12 seconds. On February 15 the cadence of EUVI-A and EUVI-B 195 {\AA} was 5 minutes. For the present study the 5 minute cadence is sufficient.

The CH boundary is determined manual by comparing the three vantage point observations, it is drawn in white in figure \ref{fig_fullsun}. We also tried using a threshold segmentation method to automatically find the CH boundary, as was done by \citet{Gopalswamy_09},  but found that the segmented boundary from one vantage point did not appropriately outline the CH as seen from a different vantage point.

To analyze the kinematics and other effects we construct stack plots. These plots are made by . To get a sense of the thermal changes in the corona during the passage of the front we compare observations from four AIA instrument channels sensitive to nominal coronal temperatures on the order of $10^6$ MK. The AIA instrument channels used in this study, from coolest to hottest peak temperature response, were AIA 171 {\AA} (Fe IX) at 0.6 MK, AIA 193 {\AA} (Fe XII) at 1.5 MK, AIA 211 {\AA} (Fe XIV) at 2.0 MK, and AIA 335 {\AA} (Fe XVI) at 2.5 MK \citep{ODwyer_10,Lemen_11}. AIA has an automatic exposure control (AEC) in cases of intense events, such as the 2011 February 15 flare, that reduces the exposure every other image but also reduces the signal-to-noise ratio \citep[see][for more discussion on AEC during this event]{Schrijver_11}. For consistency we used only AIA images that had the standard exposure time, which for AIA 171 {\AA} and 193 {\AA}  is $\sim2$ s, and 211 {\AA} and 335 {\AA} is $\sim2.9$ s. If non-standard exposure time images were included in the stack plots it would cause vertical stripes to be seen corresponding to image pairs used for running/base ratio that had different exposure times on the account of the images having different background noise distributions. Therefore, the cadence of AIA data used in this study was $\sim24$ seconds.

\section{RESULTS}\label{sec_results}

\subsection{Kinematics of the Primary Wave}\label{sec_primary}

\begin{figure}
    \centerline{\hspace*{0.0\textwidth}
   \includegraphics[width=1\textwidth]{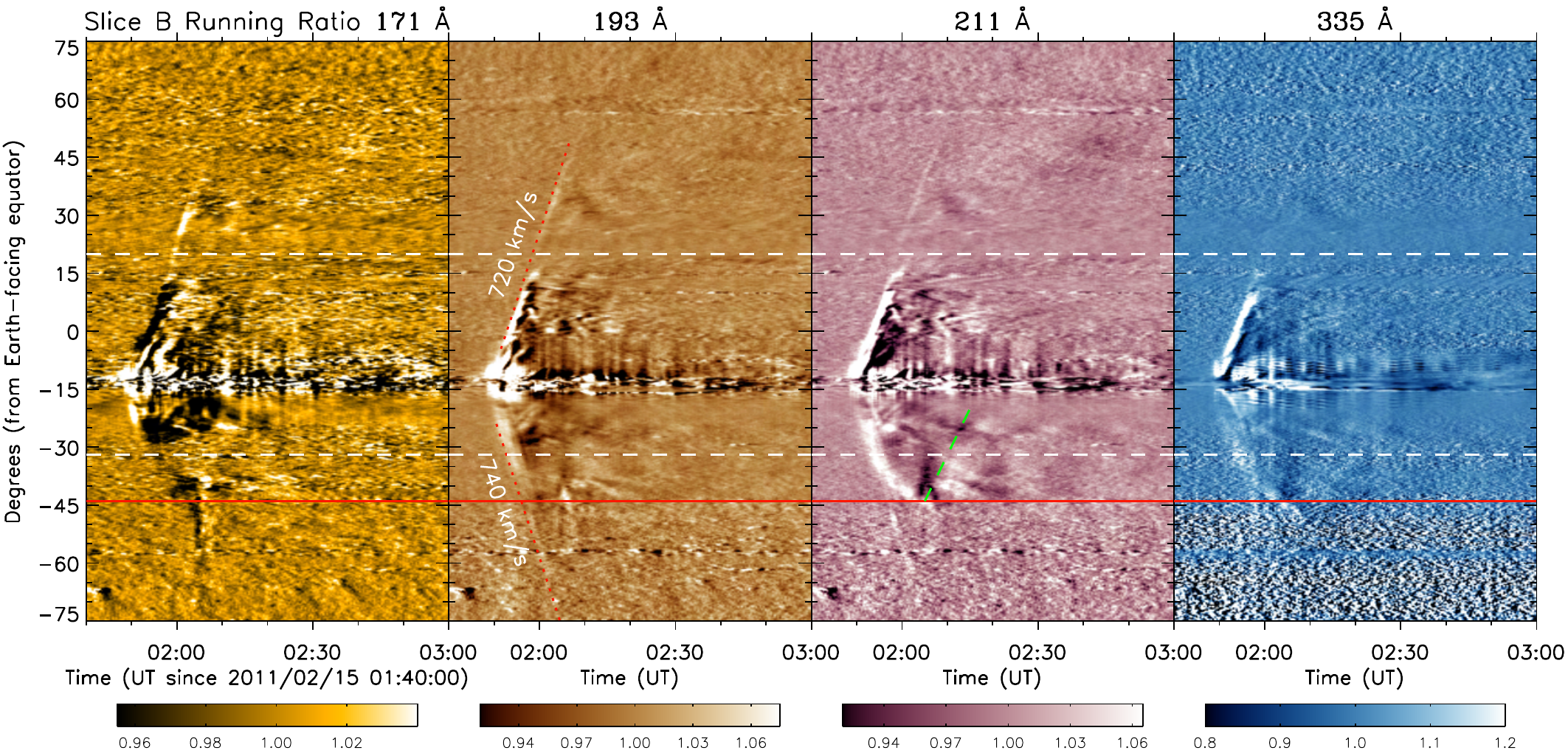}
   }
      \vspace{0.02\textwidth}
       \centerline{\hspace*{0.0\textwidth}
      \includegraphics[width=1\textwidth]{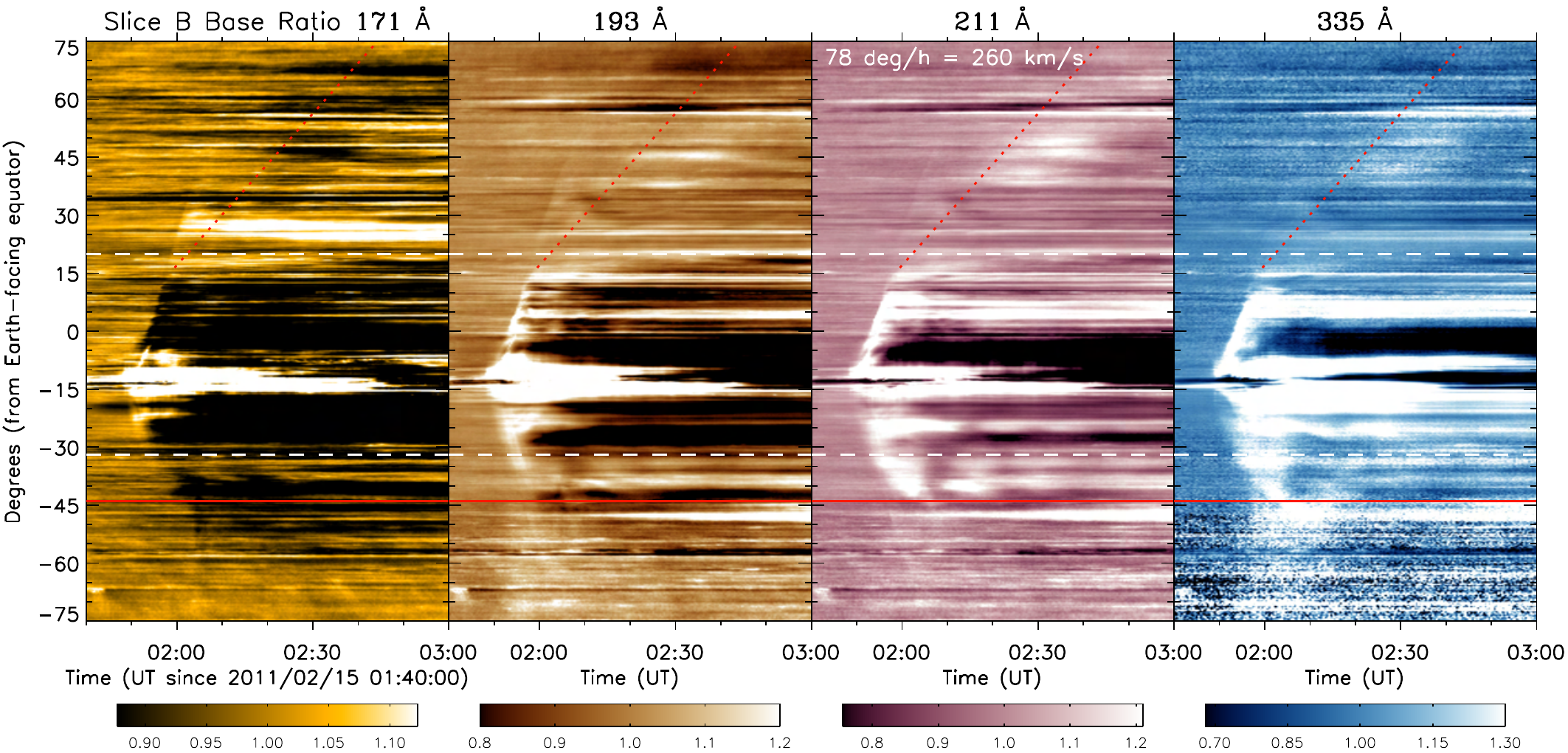}
   }
   \caption[]{Slice ``B'' that lies $9^\circ$ from the meridian at the equator. All four AIA channels are shown here. The top panel shows running ratio stack plots, and the bottom shows base ratio. This figure demonstrates the bifurcation of the wave into a fast (shown with the running ratio plots) and slow (shown with the base ratio plots) component. The horizontal red line at $-44^\circ$ is the coronal hole boundary and the two horizontal dash lines at $-32^\circ$ and $20^\circ$ are the post-CME flanks.}
   \label{fig_sliceB}
\end{figure}

\begin{figure}
    \centerline{\hspace*{0.0\textwidth}
   \includegraphics[width=0.6\textwidth]{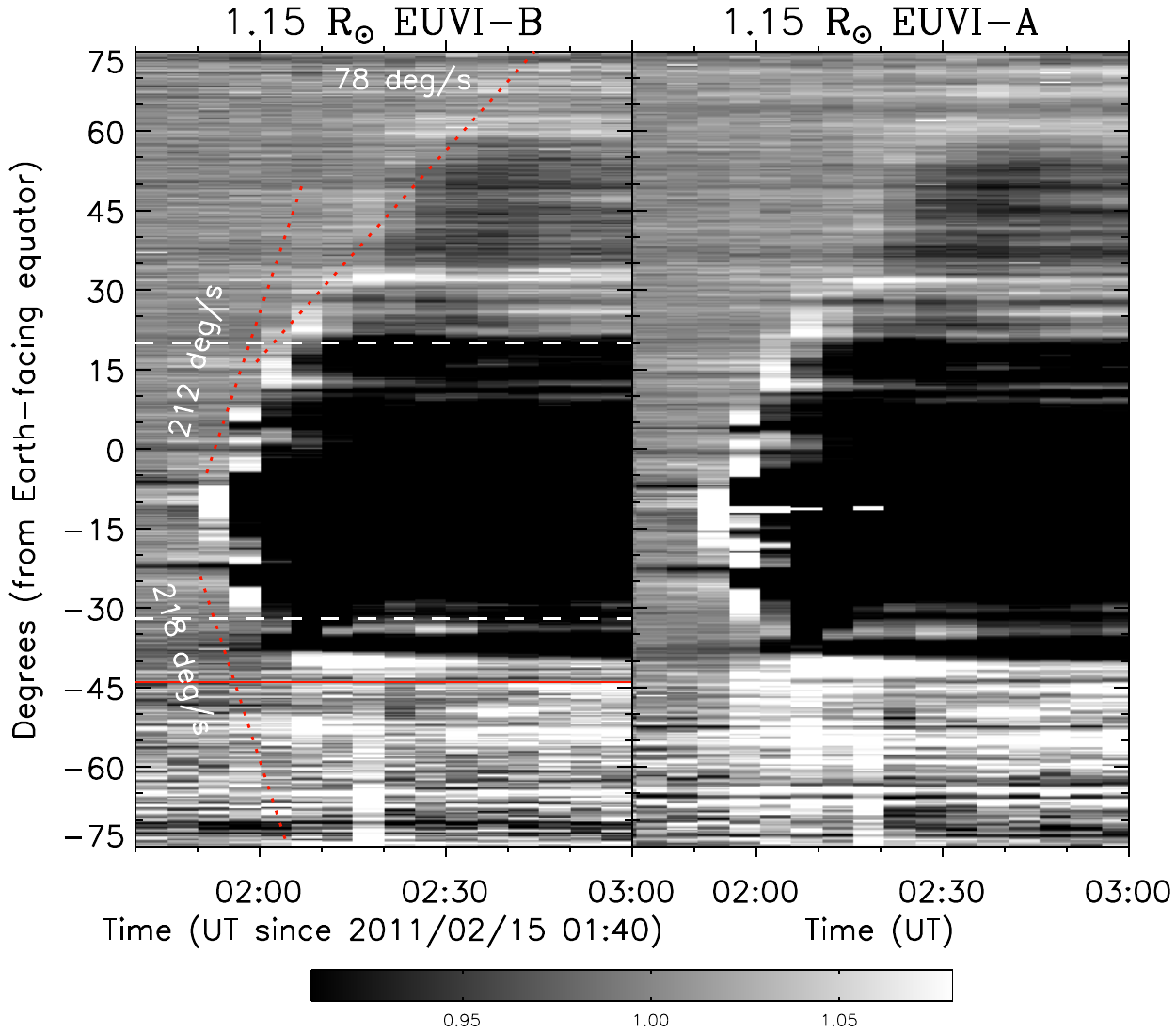}
   }
   \caption[]{Above the limb base ratio stack plot with EUVI-A 195 {\AA} and EUVI-B 195 {\AA} observations of concentric circle around the Sun at 1.15 R$_\sun$. The horizontal red line at $-44^\circ$ is the coronal hole boundary and the two horizontal dash lines at $-32^\circ$ and $20^\circ$ are the post-CME flanks.}
   \label{fig_euviab}
\end{figure}

In this section we present the measurement of the kinematics of the primary wave and compare AIA with off limb EUVI observations. To make a direct comparison we make a stack plot of a great circle ground track slice of width 36.6 arcsec off set from the meridian at the equator by $9^\circ$, labeled as ``B'' in figure \ref{fig_aia}. Figure \ref{fig_sliceB} shows the corresponding stack plot made with the four AIA channels, where the top row is 4.8 minute running ratio, and the bottom row is base ratio, with the base time at 01:35 UT. The axes are in heliographic coordinates where the origin is the Earth-facing equator, positive values are toward the north, and negative towards the south, in this way we can have a more direct comparison with the off limb EUVI observations. Figure \ref{fig_euviab} are EUVI-A and EUVI-B 195 {\AA} base ratio stack plots, with base time at 01:35 UT, of a concentric circle slice around the Sun at a radius of 1.15 R$_\sun$ and of width 50 arcsec. Figure \ref{fig_seq} middle row shows the concentric circle arc slices overlaid on the EUVI-A and EUVI-B images. We first focus on the top row of figure \ref{fig_sliceB} in which the velocity of the front is measured as it propagates towards the north and south. The velocity measured projected onto the surface towards the north was $720 \pm 20$ km s$^{-1}$ and to the south was $740 \pm 30$ km s$^{-1}$. The error was estimated from making the fit ten times by placing two points along the front and deriving the velocity. The velocity measured towards the north is in agreement with \citet{Schrijver_11} that reported $730$ km s$^{-1}$. Although the north/south velocities are approximately equal, their tracks seem to be offset in time by $\sim2$ minutes with the south ahead of the north. This asymmetry could be due a projection effect, considering that the southern track is closer to the limb. The red horizontal line at $-44^\circ$ is the CH boundary towards the south of the flare site. An interesting feature in this stack plot is an apparent transmission of the wave front through the CH propagating at the same velocity as the primary wave. This feature is most evidently seen in AIA 193 {\AA} and AIA 211 {\AA} observations. As can be seen in the AIA 193 {\AA} running ratio movie (online) the transmission occurs all along the CH boundary. Though we must be cautious because it may be that what we are seeing is due to a projection effect of the wave deflected above the CH. We return to this topic in section \ref{sec_transmit}.

The two dashed horizontal lines in figure \ref{fig_sliceB} are positions identified by \citet{Schrijver_11} as the location of the post-CME flanks, which we interpret as the angular extent of the CME. In the work of \citet{Schrijver_11} the velocity reported to the north is fitted only up to the northern post-CME flank at $20^\circ$. We note that between $\sim15^\circ$ and $\sim40^\circ$ the slice passes through part of an active region. However, our extended plot reveals bifurcation of the front into fast and slow components. The fast component can be seen in both running ratio and base ratio stack plots in figure \ref{fig_sliceB}. The front, as observed in AIA 171 {\AA} is initially seen as a dimming and an enhancement in the other three AIA channels. This observation can be explained as an adiabatically heated plasma due to a compression front \citep[e.g.][]{Downs_12}. On the other hand, beyond $20^\circ$ the front is observed as an enhancement in the four AIA channels. This is interpreted as maybe due to either a density enhancement and/or a transverse/kink mode wave traveling along the loops and not a temperature effect. (see AIA 193 {\AA} movie).

The slow component is identified in the base ratio stack plots and is most prominently seen in AIA 211 {\AA}. The front is faintly observed in AIA 193 and 335 {\AA}, and faintly observed as a dimming in AIA 171 {\AA} indicating adiabatic heating. We attempt to measure the velocity and find a value of $260 \pm 20$ km s$^{-1}$. We compare AIA observations with off limb EUVI observations in Figure~\ref{fig_euviab}. The EUVI-A/B stack plots look essentially identical because they are observing almost exactly $90^\circ$ from the Earth-facing disk center in opposite directions. Again the horizontal red line at $-44^\circ$ is the location of the CH boundary and the two dashed horizontal lines are the post-CME flanks. The velocity fits for the fast and slow components found with AIA are over-plotted onto the EUVI plots (dotted red lines) and appear reasonable considering the lower cadence of EUVI. The values written on the plot have been converted to degrees per hour, where $260$ km s$^{-1}=78^\circ$ h$^{-1}$ is the slow component,  $730$ km s$^{-1}=212^\circ$ h$^{-1}$ is the fast component towards the north, and $740$ km s$^{-1}=218^\circ$ h$^{-1}$ is the velocity toward the south. The region below the slow component shows a weak dimming in both EUVI 195 {\AA} stack plots. The intensity of the dimming is much smaller than the dimming caused by the evacuation of material under the main body of the CME (within the dashed lines in the figure). The relatively long duration (at least 40 minutes) of the dimming, however, suggests that it might be the result of mass evacuation rather than mass displacement. Therefore, the lateral CME expansion, and consequent coronal mass outflow, do not stop at $\sim 20^\circ$ north of the flaring site but continue northward to $\sim 60^\circ$ but at a much diminished rate. An alternative explanation could be that after $\sim 20^\circ$, the location of the CME flank, a fast-mode wave is launched at the local fast-mode speed of $260$ km s$^{-1}$ and that the dimming in EUVI is due to heating of the tenuous plasma. This is a plausible explanation considering that this velocity is comparable to the fast-mode speeds measured in other parts of the Sun associated with this event. It is also consistent with the hybrid interpretation of coronal waves because the slow component could be the fast-mode wave left traveling close to the solar surface that is no longer being driven by the CME, which has propagated radially away. The fact that the AIA observations show an enhancement and EUVI show a dimming along that track can be understood as a line of sight effect. The AIA telescopes look straight down the out flowing material while in the EUVI imagers look above the limb. In any case, the dimming region in the EUVI stack plots exactly corresponds to the same region of enhancement in AIA indicating that we are tracking the same feature. Note that this feature is not observed towards the south where the CH lies. Obviously the CME cannot expand into the CH and hence no outflowing material is observed, on the other hand any heating by the fast-mode wave traveling through the CH may not be observed because of the low density there.

%%%%%%%%%%%%%%%%%%%%%%%%%%%%%%%%%%%%%%%%%%%%%%%%%

\subsection{Kinematics of the ``Reflected Wave'' Front}\label{sec_reflection}

\begin{figure}
   \vspace{-0.1\textwidth}
   \centerline{\hspace*{0.0\textwidth}
   \includegraphics[width=0.267\textwidth]{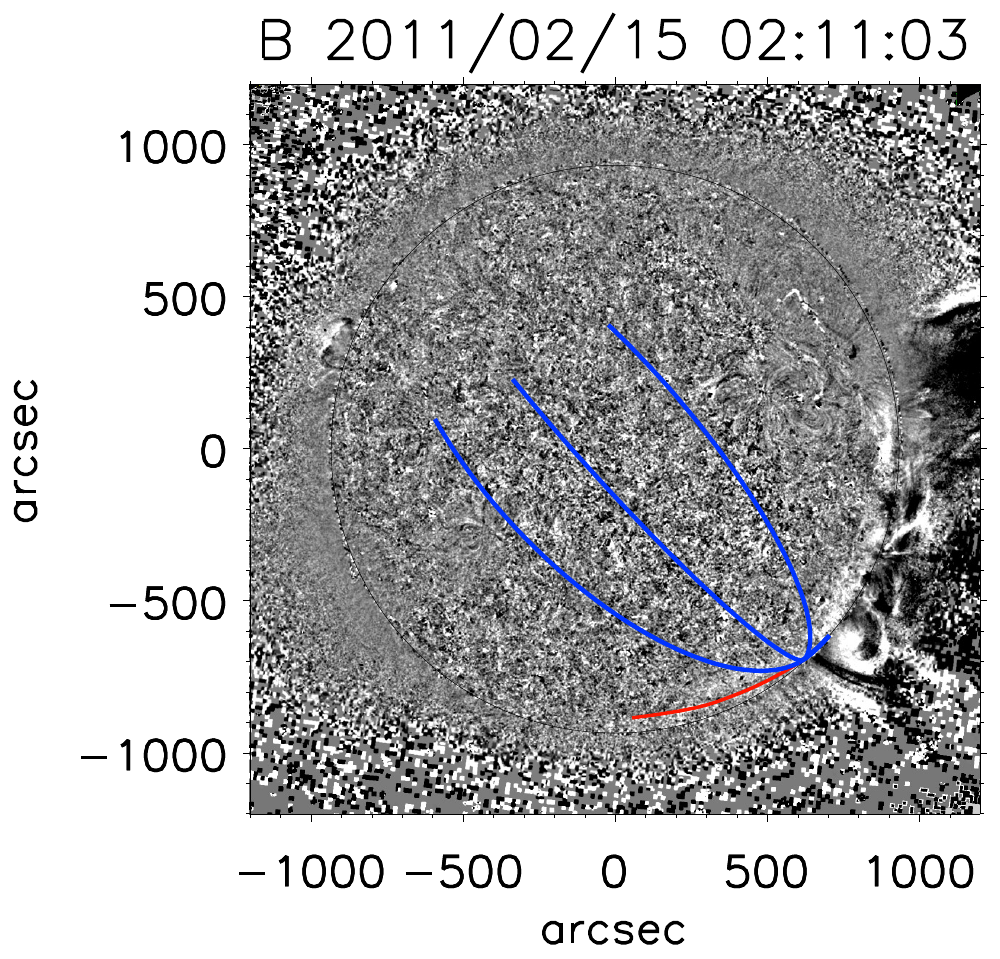}
              \includegraphics[width=0.466\textwidth]{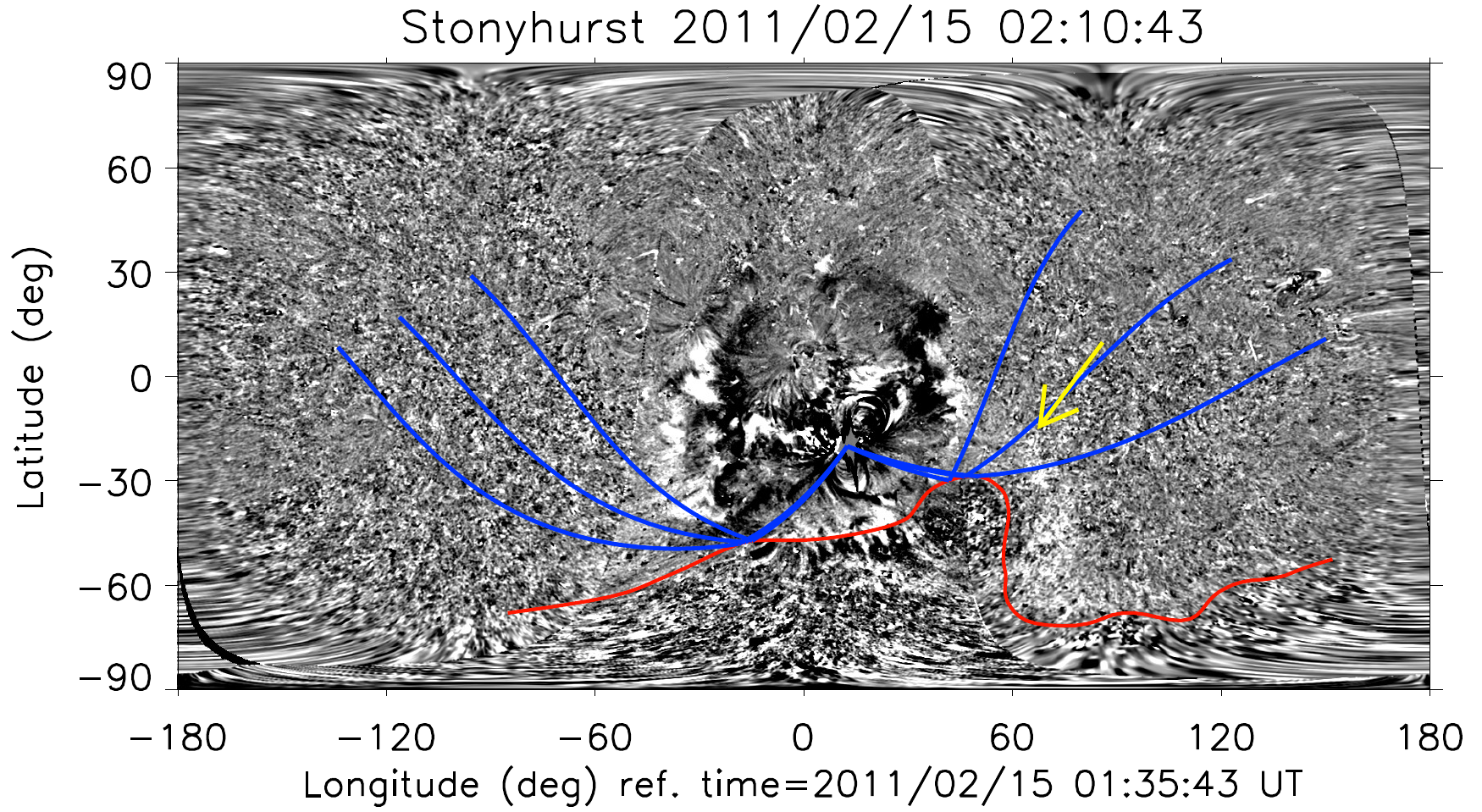}
                            \includegraphics[width=0.267\textwidth]{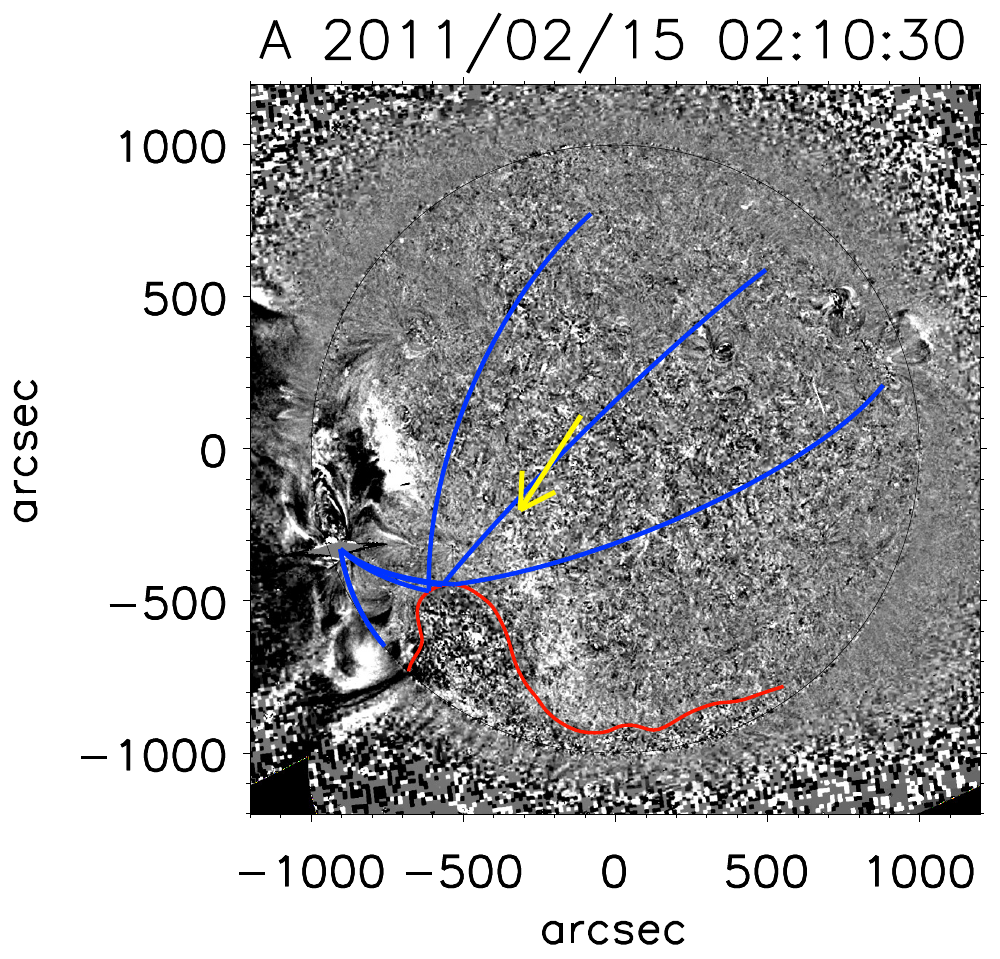}
   }
   \vspace{0.01\textwidth}
   \centerline{\hspace*{0.0\textwidth}
   \includegraphics[width=0.267\textwidth]{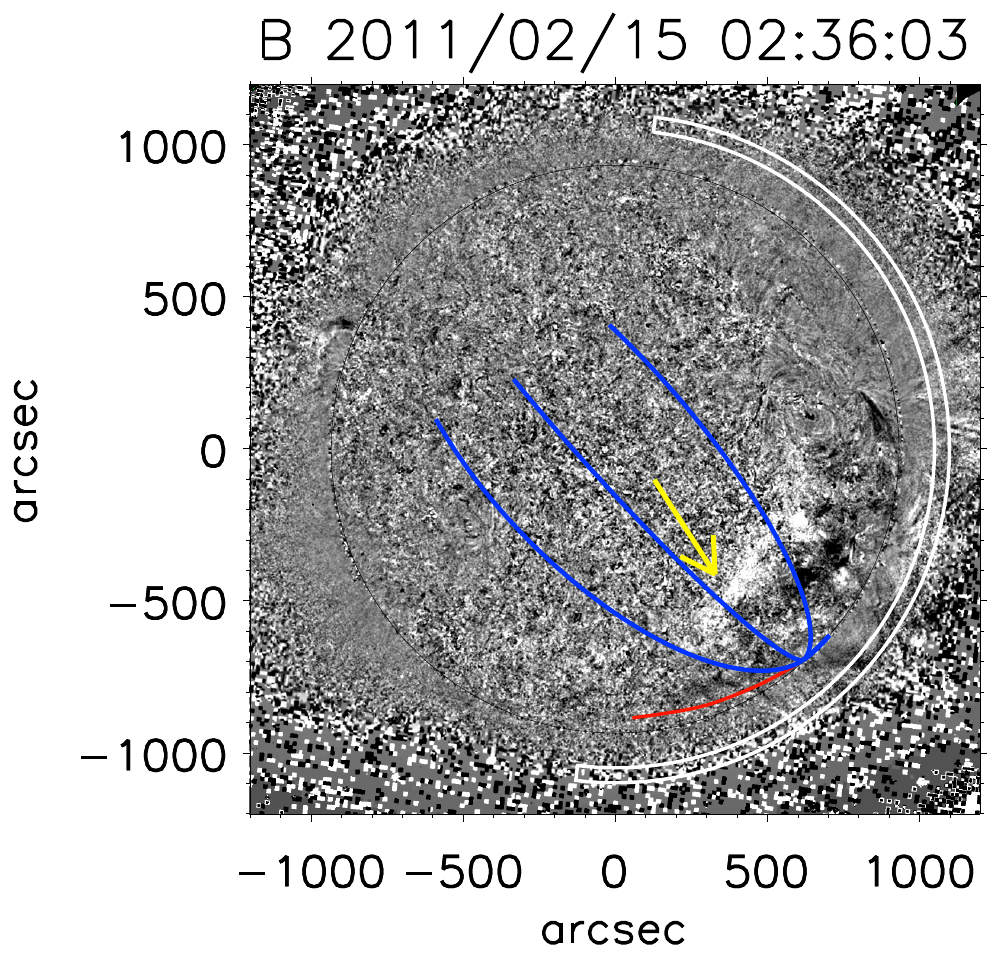}
              \includegraphics[width=0.466\textwidth]{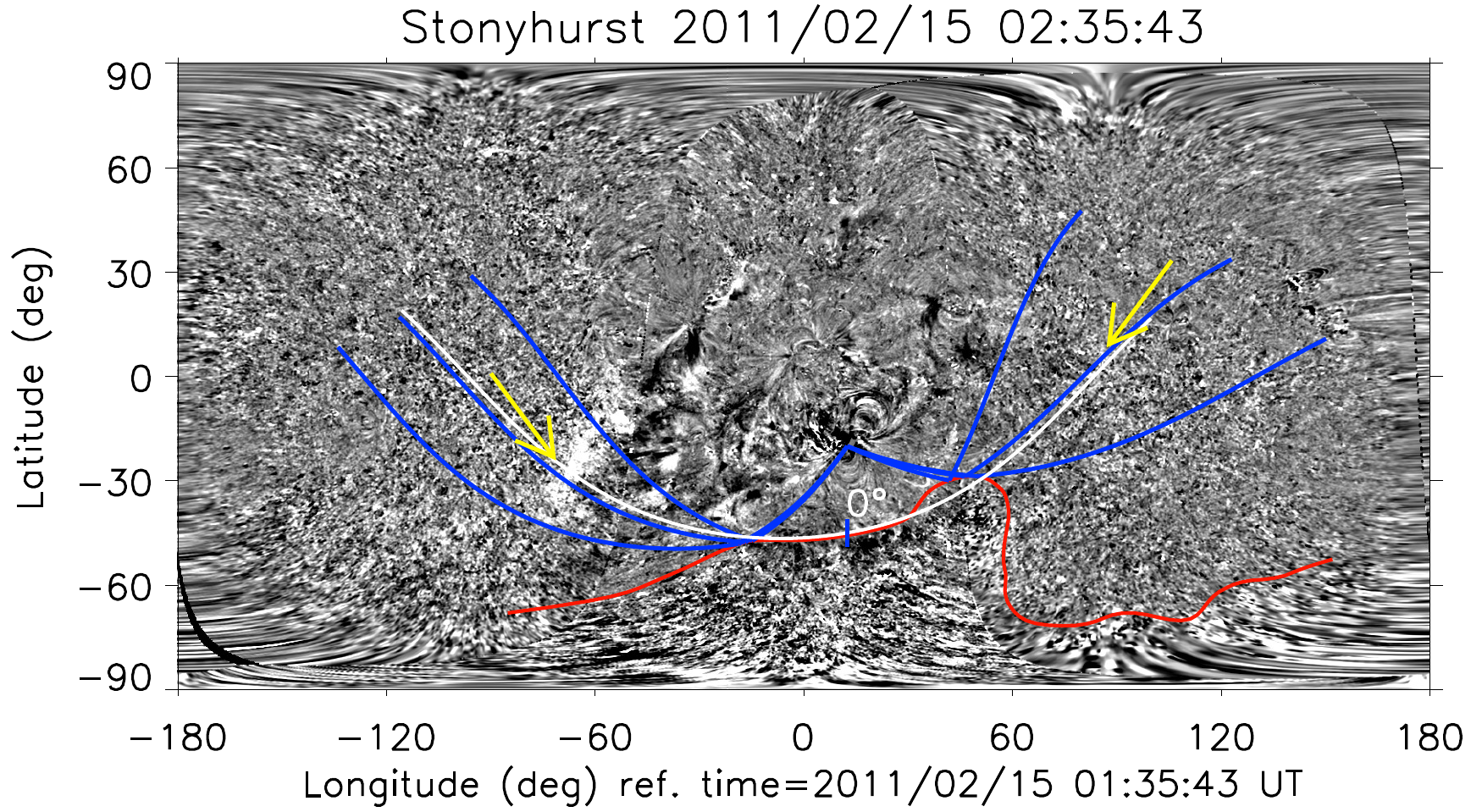}
                            \includegraphics[width=0.267\textwidth]{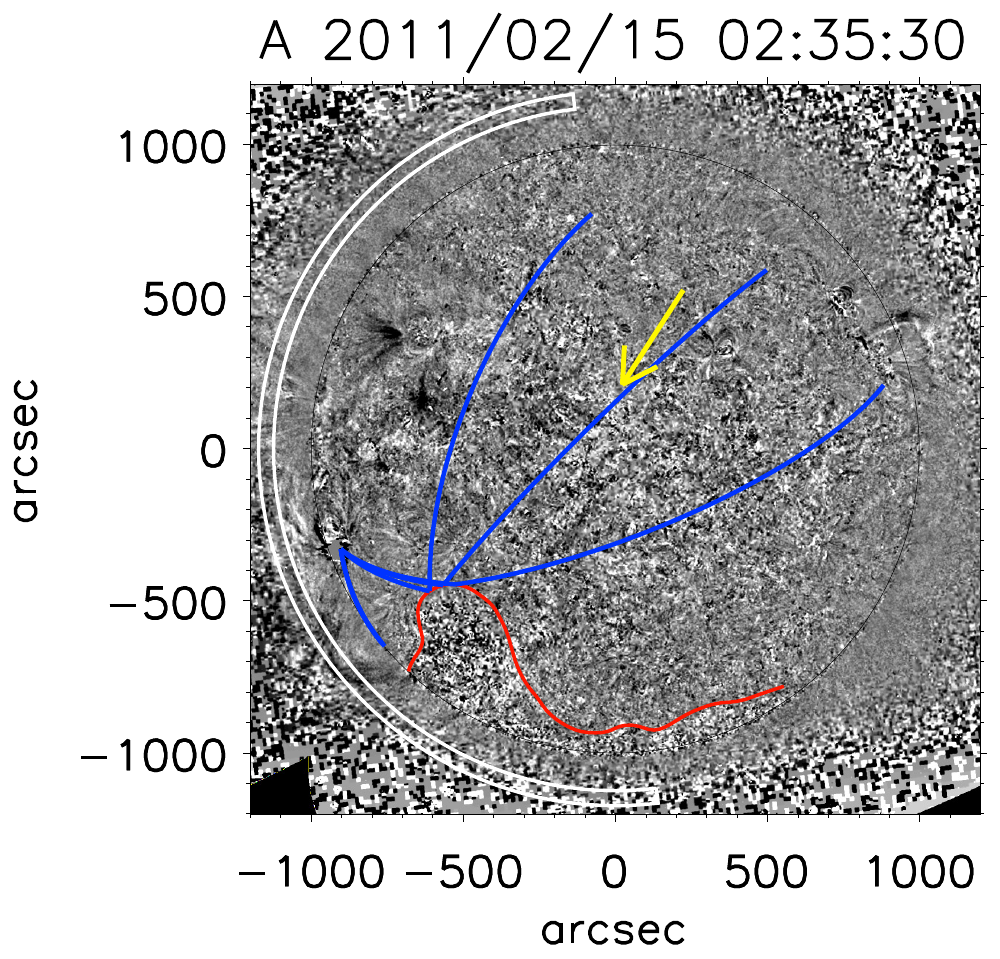}
   }
   \vspace{0.01\textwidth}
         \centerline{\hspace*{0.0\textwidth}
   \includegraphics[width=0.267\textwidth]{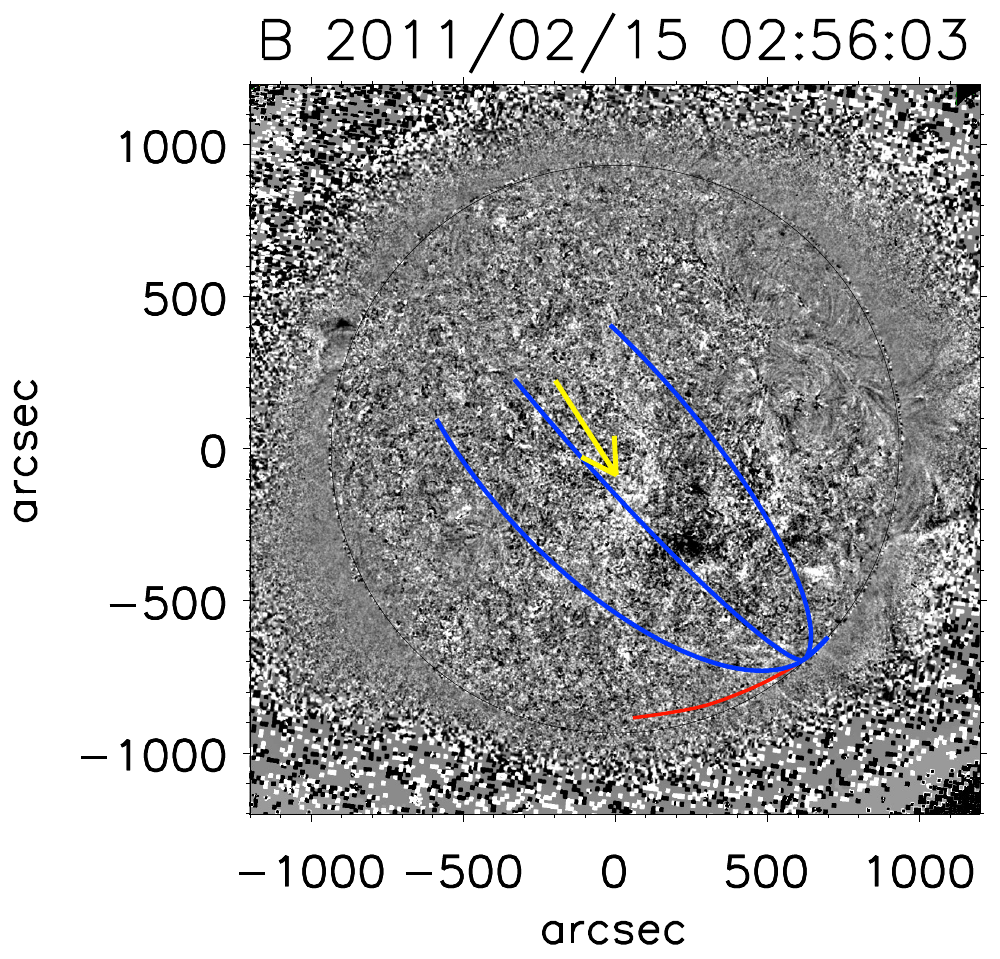}
              \includegraphics[width=0.466\textwidth]{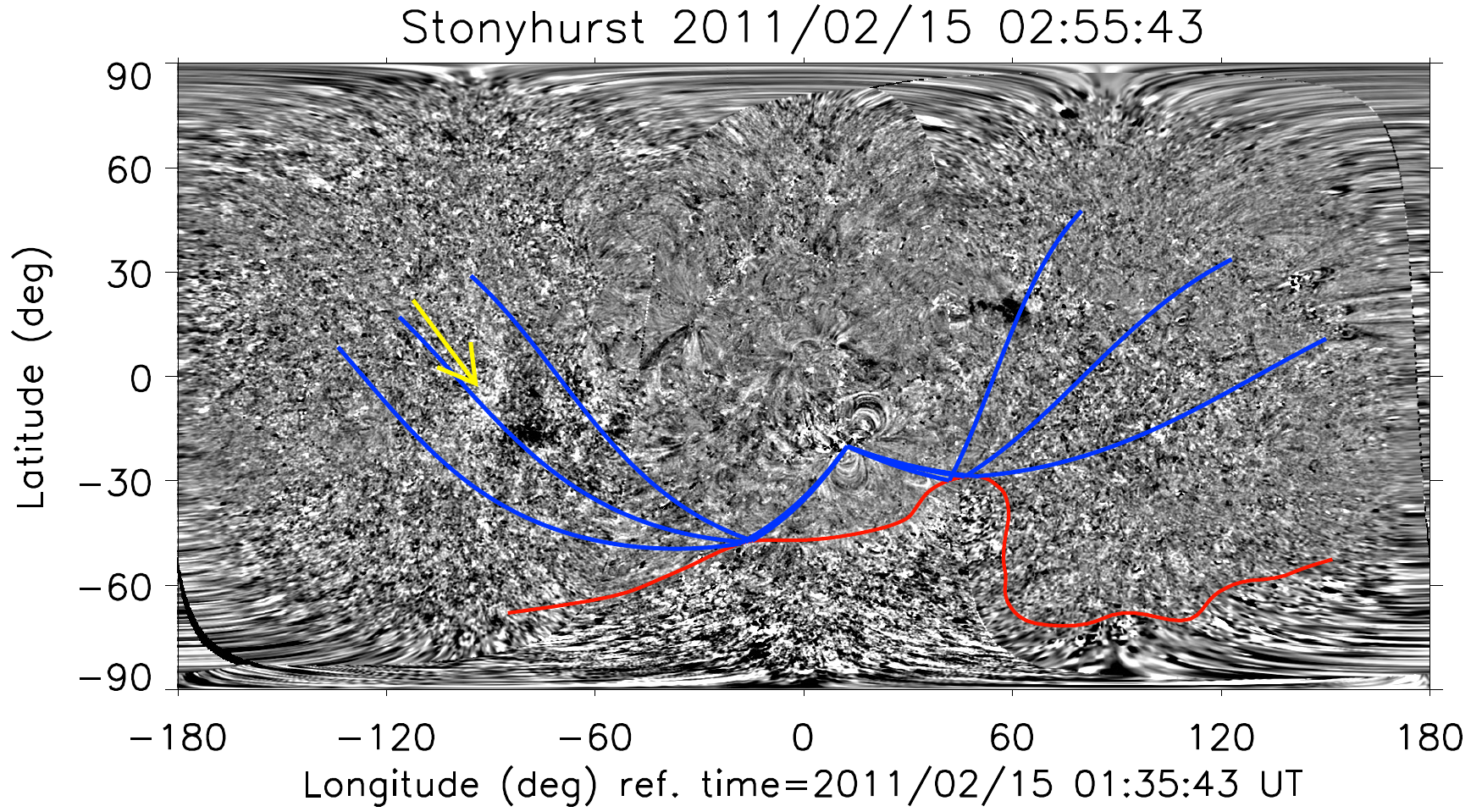}
                            \includegraphics[width=0.267\textwidth]{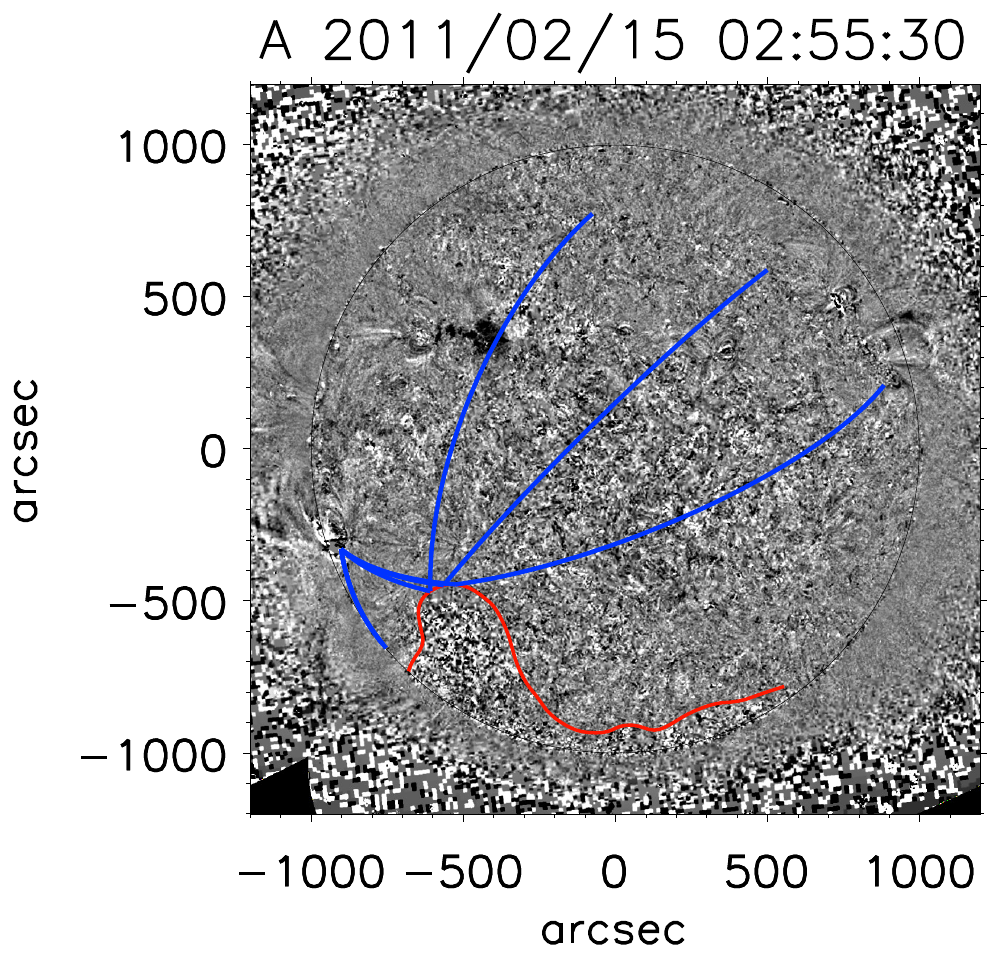}
   }
   \caption[Sequence of Full Sun Maps]{Sequence of running ratio frames showing the evolution of the coronal wave at different time steps. Left column is EUVI-B 195 {\AA}, the right column is EUVI-A 195 {\AA}, and the center column is the full Sun map that combine EUVI-A and EUVI-B with AIA 193 {\AA}. Yellow arrows indicate the wave front. Blue lines are the predicted trajectory, indicating the reflection from the coronal hole boundary. The red line is the coronal hole outline. The white line drawn on the full Sun map of 02:35 UT (center) is slice A. See movies online.}
   \label{fig_seq}
\end{figure}

We have applied the technique of making full Sun maps, as described in section \ref{sec_obs}, to a sequence of images from 01:30 to 03:35 UT with a cadence of 5 minutes. Running ratio images were compared with base ratio images (using a base image at 01:35 UT) and it was found that utilizing either technique does not change our results. Figure \ref{fig_seq} shows a sample of running ratio images from the sequence. Movies of the sequence (full Sun map as well as AIA 193 {\AA}, EUVI-A and EUVI-B) are provided online. The left column is EUVI-B, the center is the map, and the right is EUVI-A. The CH boundary (red) is projected onto the FOV of each image.

Since we are interested in the possibility of wave ``reflection'' from CH boundaries we perform a ray trace analysis. This is strictly for visualization purposes to guide the eye to the motion of the propagating front. We consider six trajectories originating from the flare site, following great circle paths, and terminating at the CH boundary. Three of the paths extend to the east and three to the west of the source region. The incident and reflected angle is calculated with respect to the smoothed CH boundary. Then the six trajectories are extended along the reflected path. Note that these are our predicted paths for the trajectory of the coronal wave obeying the law of reflection purely under the assumption that the incident and reflected angles are equal. The eastern reflection is more easily observed than the western one, which appears more diffused. After $\sim$ 2:45 UT the western reflection becomes too diffuse to be seen. The eastern reflection is visible until $\sim$ 3:20 UT. Looking carefully at the rays drawn one notices that initially the three rays to the East and to the West appear to almost overlap, yet in the FOV of the EUVI-A and EUVI-B the reflected rays diverge greatly. This is due to convexity of the CH boundary and is sensitive to how that boundary is drawn. In any case, the rays drawn were carefully chosen by adjusting the incident angle and the position at the CH boundary where the reflection takes place to emphasize the trajectory of the front within the FOV of EUVI-A and EUVI-B.

\begin{figure}
    \centerline{\hspace*{0.0\textwidth}
   \includegraphics[width=0.6\textwidth]{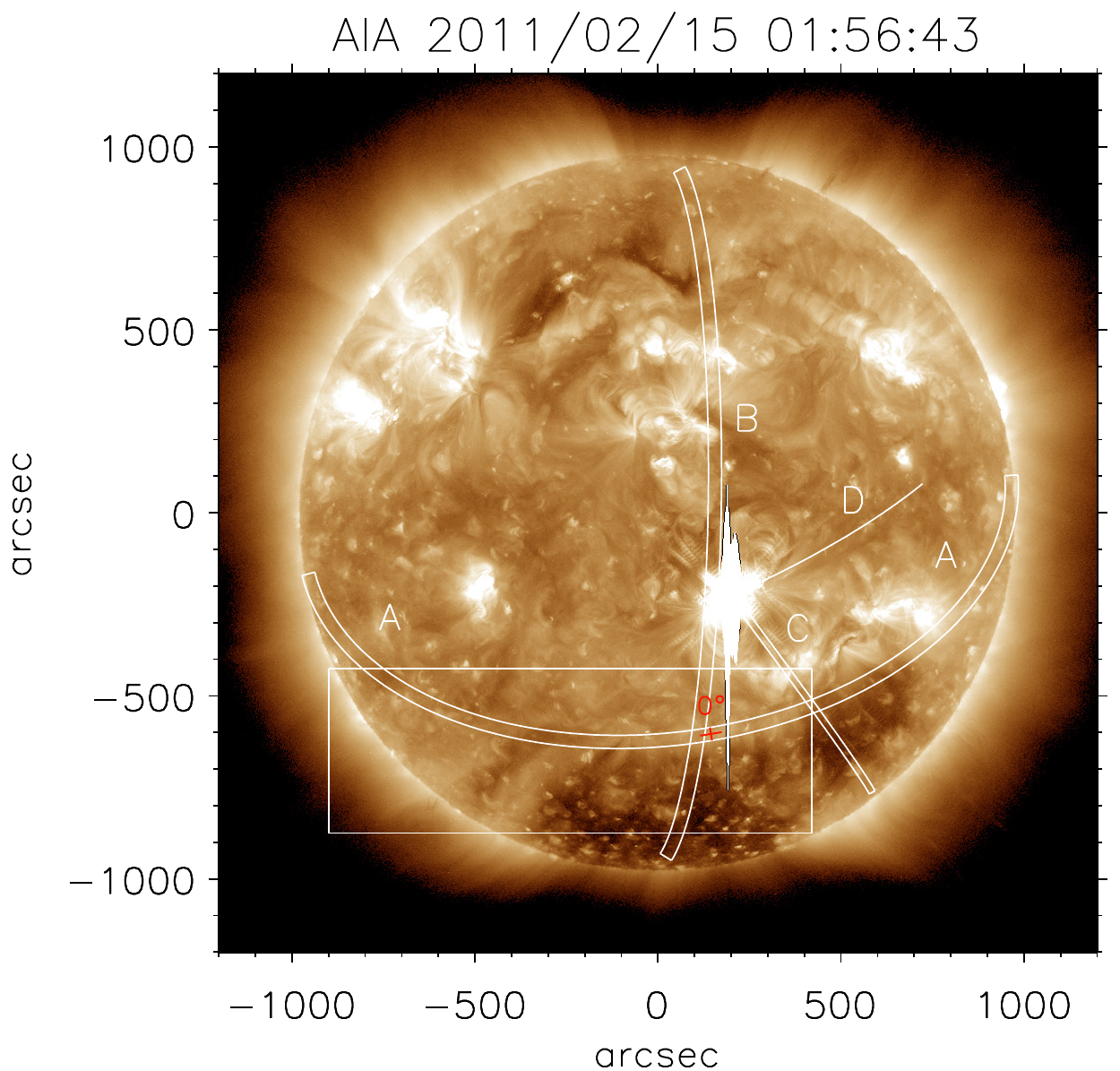}
   }
   \caption[]{SDO/AIA 193 {\AA} image showing the ground track slices made for this study. Note that the tracks follow great circles. The location of zero degrees along slice ``A'' is marked by a red ``+''.}
   \label{fig_aia}
\end{figure}

\begin{figure}
    \centerline{\hspace*{0.0\textwidth}
   \includegraphics[width=0.5\textwidth]{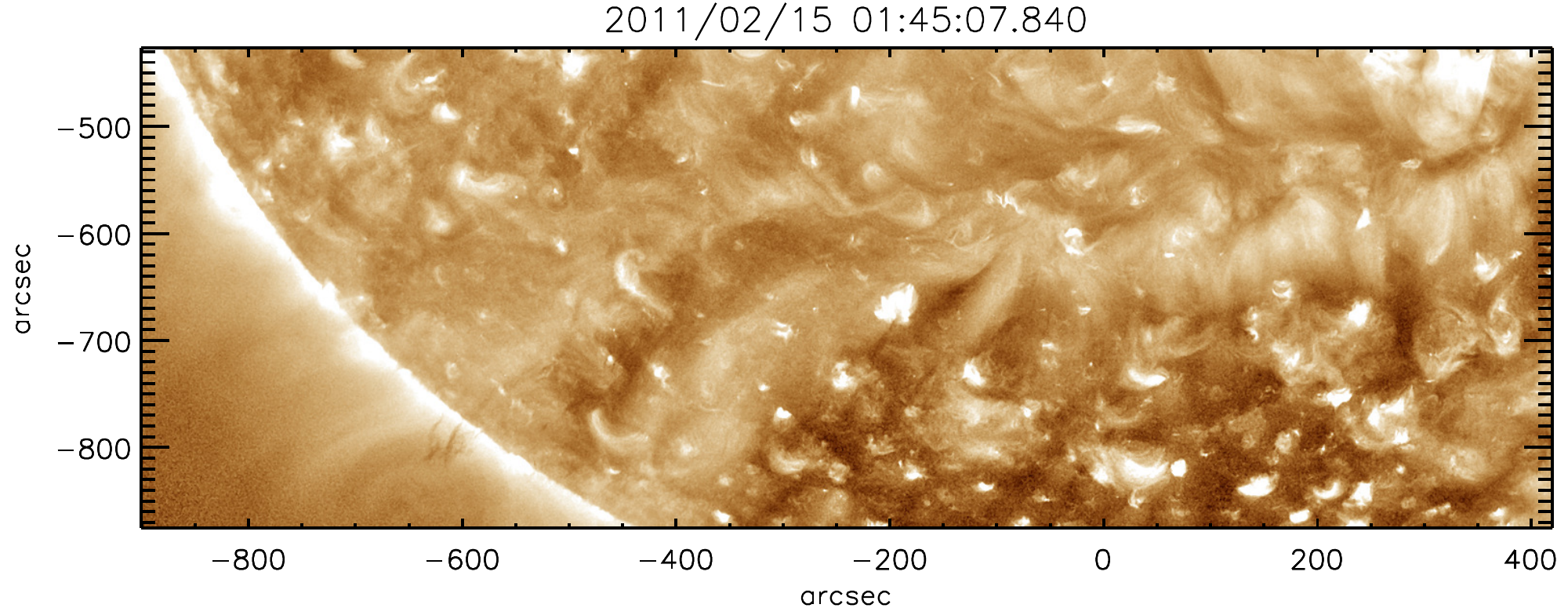}\includegraphics[width=0.5\textwidth]{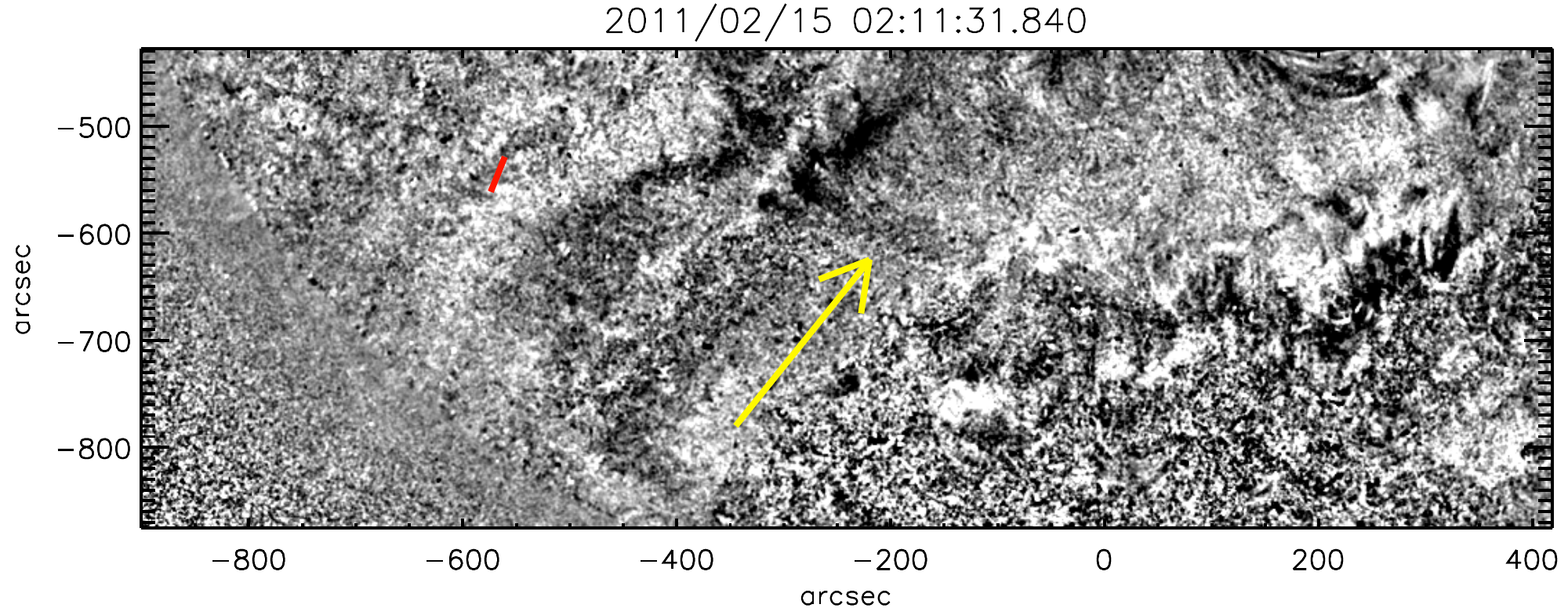}
   }
       \centerline{\hspace*{0.0\textwidth}
   \includegraphics[width=0.5\textwidth]{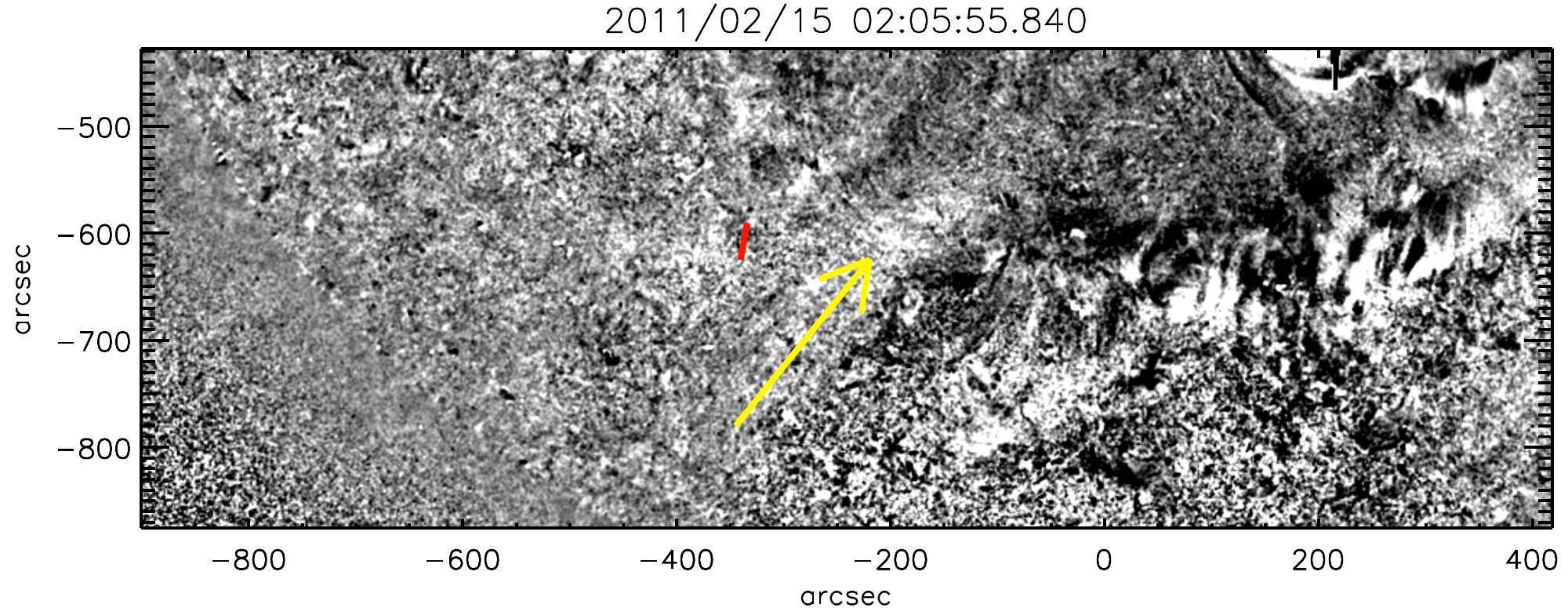}\includegraphics[width=0.5\textwidth]{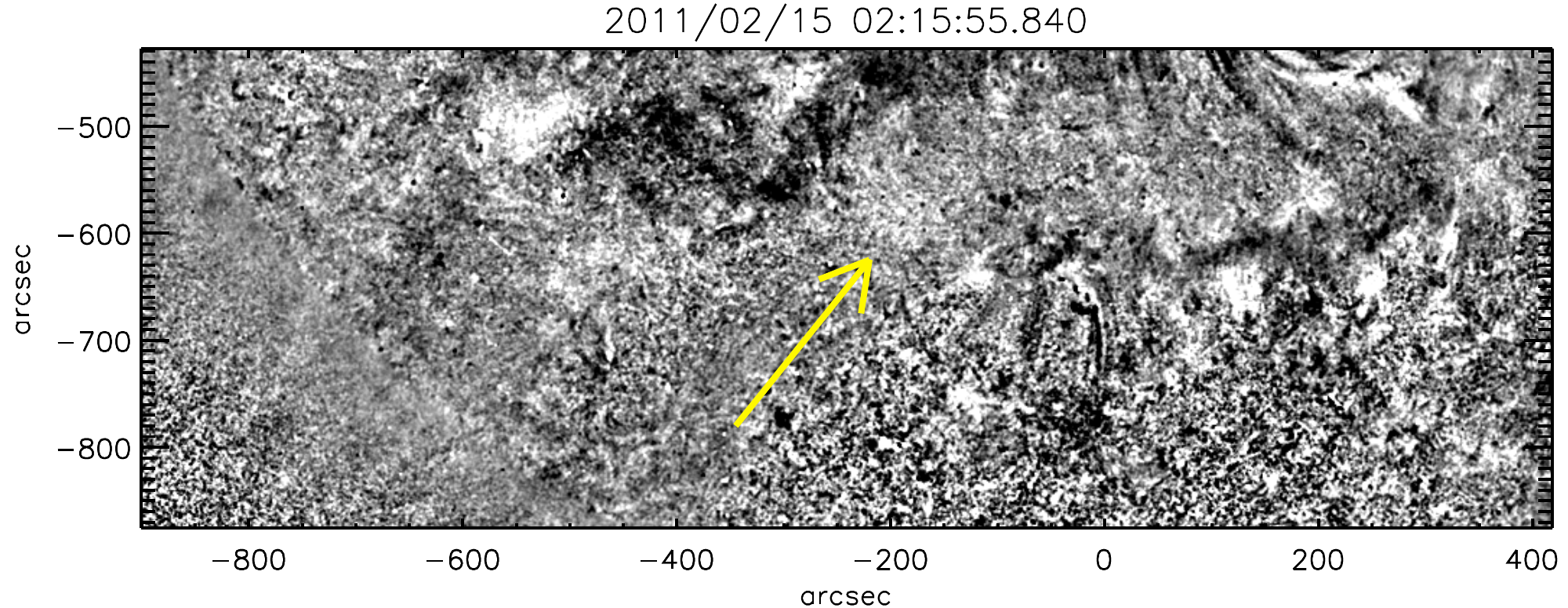}
   }
   \caption[]{Cutout region, indicated by the rectangle in figure \ref{fig_aia} showing the loops at the coronal hole boundary. Top left panel is an image enhanced using a wavelet technique. The arrows point to secondary waves apparently launched by the oscillating loop arcade. The red mark at 02:05 UT and 02:11 UT indicates the position of the wave front as seen along slice ``A'' (see figure \ref{fig_sliceAzoom}). See movies online.} 
      \label{fig_cutout}
\end{figure}

\begin{figure}
    \centerline{\hspace*{0.0\textwidth}
   \includegraphics[width=0.6\textwidth]{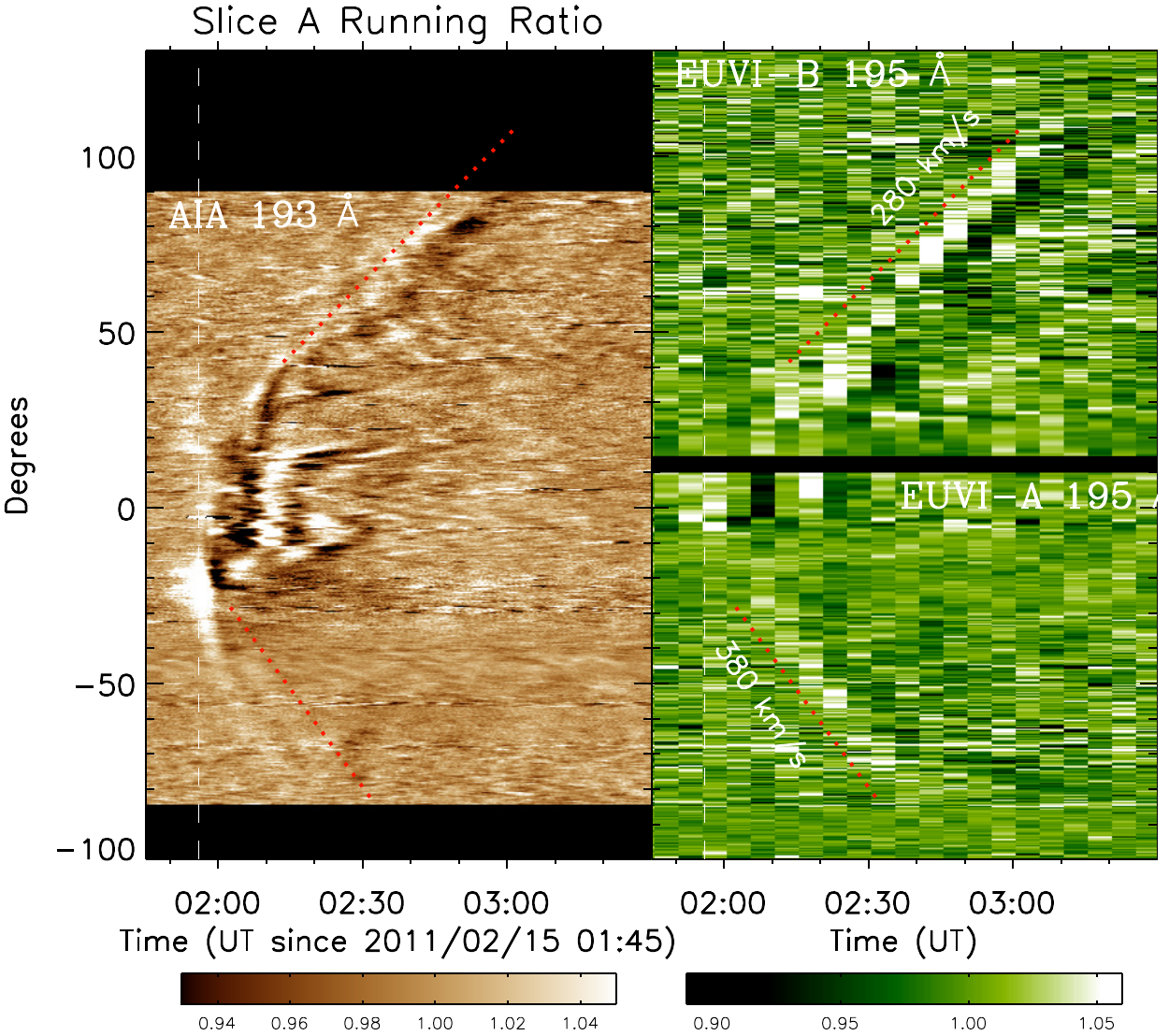}
   }
   \caption[]{Ground track Slice ``A'' that spans $230^\circ$ across the Sun. On the left is the AIA 193 {\AA} stack plot and on right are the EUVI-A 195 {\AA} (bottom portion) and EUVI-B 195 {\AA} (top portion) stack plots.}
   \label{fig_sliceA}
\end{figure}

Next we placed a great circle path chosen to follow the trajectory of the wave front in the FOV of EUVI-B and extended it through the FOV of AIA and EUVI-A. This path was made independently from the ray tracing analysis described. It was coincidentally found that this path was aligned nearly parallel with the reflected ray trace in both the EUVI-A and EUVI-B FOV (figure \ref{fig_seq} full Sun map middle panel) for this reason we call this the ``reflected'' wave. Another interesting find is that this path is aligned with the CH boundary. Close inspection reveals the existence of a loop arcade straddling this part of the CH boundary, a cutout of the loop region is shown in figure \ref{fig_cutout}. The kinematics of the wave and the role of the loops are further studied by making stack plots using the AIA and EUVI instruments along this great circle path. This stack plot is shown in figure \ref{fig_sliceA}, and is directly comparable to figure \ref{fig_seq}. The left panel is AIA 193 {\AA} 4.8 minutes running ratio, and the right panel shows both EUVI-A and EUVI-B with 5 minutes running ratio. Here we have removed the seam and instead show all of the available spatial observation to directly compare what is seen within the FOV of the three instruments. The width of the slice used to make the stack plot in the AIA FOV was 36.6 arcsec, shown in figure \ref{fig_aia} labeled as ``A'', and the width of the slice in the EUVI FOV was 50 arcsec. Zero degrees in figure \ref{fig_sliceA} is marked in figure \ref{fig_seq} middle panel, and \ref{fig_aia}. The full angular extent of the wave into the FOV of EUVI-A and EUVI-B along this path is $\gtrsim200^\circ$ (figure \ref{fig_sliceA} right panel). If measured from the flare site, the extent to the west (into the FOV of EUVI-A) is $\sim80^\circ$ and to the east (into the FOV of EUVI-B) is $\sim120^\circ$, therefore the angular extent as measured from the flare site is also $\gtrsim200^\circ$.

Further out in the FOV of EUVI-A and EUVI-B (figure \ref{fig_sliceA} right panel) the velocity of the ``reflected'' wave appears to be linear. To the east, into the FOV of EUVI-B the velocity was found to be $280 \pm 10$ km s$^{-1}$, and to the west into the FOV of EUVI-A, $380 \pm 25$ km s$^{-1}$. Again, the error was estimated from making the fit ten times by placing two points along the front and deriving the velocity. These values resemble those found by \citet{Schrijver_11} ($270 \pm 35$ km s$^{-1}$ and $340 \pm 35$ km s$^{-1}$ to the east and west respectively). Though it should be noted that they just estimated the velocity by finding the Carrington longitudinal position of the front at 03:00 UT ($-60^\circ$ and $+105^\circ$ to the east and west respectively) and estimated the velocity to reach that longitude from the flare site longitude.

Comparing what is observed in AIA versus EUVI-A we see that the fit in the FOV of EUVI-A does not exactly correspond to what is seen in AIA, unlike between AIA and EUVI-B. A plausible explanation is that the wave front (or part of wave front) observed is at some height in the corona and not close to the surface, and so will appear different depending on the perspective. Alternatively, in the FOV of AIA slice ``A'' is overlaid over an active region loop system, with loops that are projecting into the corona, and as the primary wave interacts with these loops intensity enhancement in running ratio images could be due to oscillations and/or mass motions occurring not close to the surface but at a different height along the loop. On the other hand, between AIA and EUVI-B the wave passes over quiet Sun where there is no active region or loop system. It is known that observations of a coronal wave from multiple view-points should look alike as the wave propagates away from its origin \citep{Ma_09} given it does not interact with other coronal structures. This is an indication that the wave observed is close to the solar surface and that what we see between AIA and EUVI-B is the same feature.

\begin{figure}
    \centerline{\hspace*{0.0\textwidth}
   \includegraphics[width=1\textwidth]{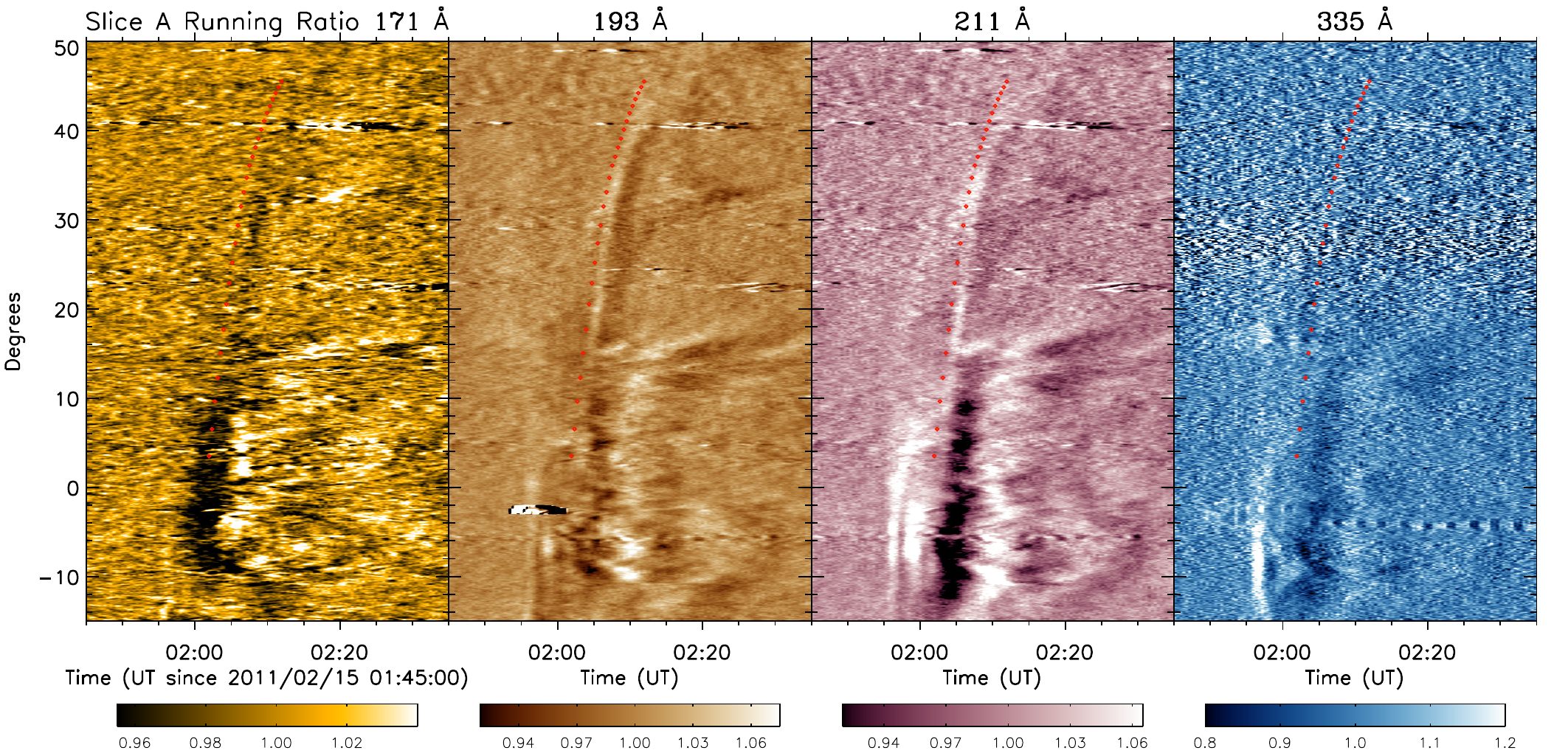}
   }
   \caption[]{Close up view of slice ``A'' stack plot in all four AIA channels to more clearly show the secondary wave triggered by the cascading loops. The measured points are indicated by the red dots.}
   \label{fig_sliceAzoom}
\end{figure}

\begin{figure}
    \centerline{\hspace*{0.0\textwidth}
   \includegraphics[width=0.6\textwidth]{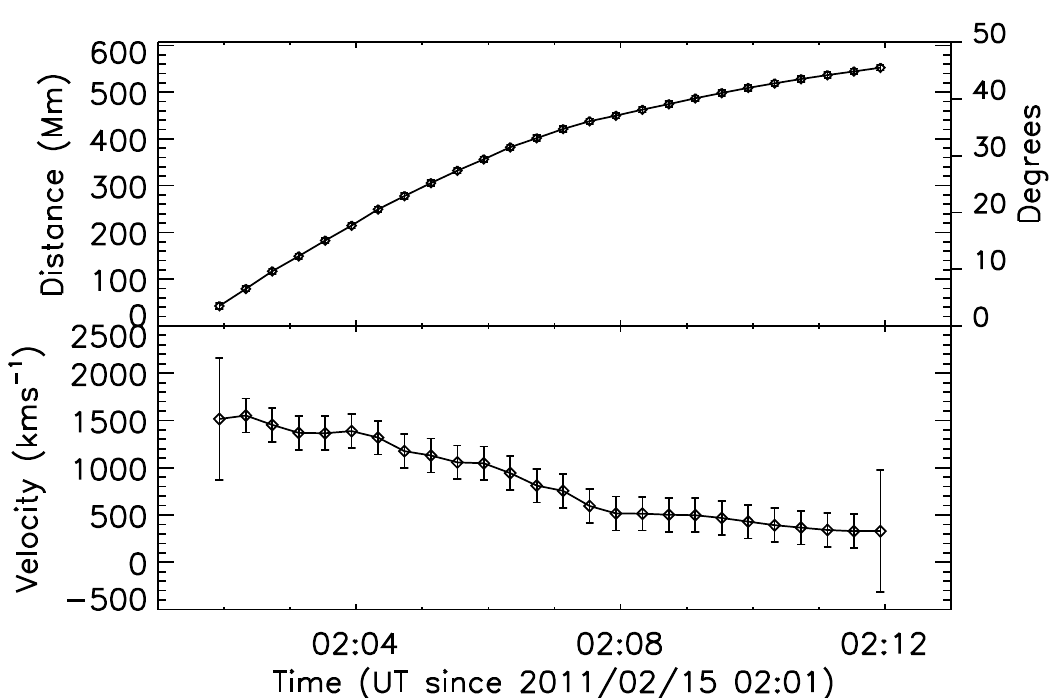}
   }
   \caption[]{Distance-time and velocity plots of the triggered wave. An error of $0.5^\circ$ was estimated for the distance measurement.}
   \label{fig_trigger}
\end{figure}

Next, we look into the loop arcade that straddles the CH boundary, shown in the cutout in figure \ref{fig_cutout} (top left) after being enhanced with a wavelet method \citep{Stenborg_03}. The other panels of figure \ref{fig_cutout} are a sequence of 1.2 minute running ratio images. The enhanced movie (provided online) of the loop arcade reveals the response of the arcade to the primary wave. The loops within the arcade are seen to oscillate perpendicular to the primary wave, then quickly damp after about two periods. It is difficult to distinguish if the loops are oscillating transversely or with the kink-mode or a combination of both as no individual loop can be identified within the arcade. Furthermore, the loop arcade is seen close to the central meridian so only the tops of the loops are observed. A similar case of a loop excited to oscillation by a primary fast-mode wave has recently been reported using SDO/AIA observations by \citet{Aschwanden_11}. The cascading and oscillating loop arcade can be seen in the AIA stack plot of figure \ref{fig_sliceA} (left panel) between $-15^\circ$ and $18^\circ$. We estimate the period to be about 10 minutes, which is consistent with measured periods of transverse loop oscillations \citep[e.g.][]{White_12}. Figure \ref{fig_sliceAzoom} shows a close up of the area between $-15^\circ$ and $50^\circ$ of figure \ref{fig_sliceA} using 1.2 minutes instead of 4.8 minutes running ratio. Using AIA 193 {\AA} and 211 {\AA} we manually track the wave front feature that extends beyond the loop arcade. We interpret this feature as a wave triggered by the cascading loop arcade, which propagates towards EUVI-B (also see the AIA 193 {\AA} running ratio movie). We propose that this is the origin of the wave which we are calling the ``reflected'' wave. Figure \ref{fig_trigger} shows the kinematics of this wave, we approximate a conservative measure error of $0.5^\circ$. From $0^\circ$ to $18^\circ$ or from 02:02 UT to 02:04 UT the speed has an average value of $\sim1400$ km s$^{-1}$. This is the velocity at which the loop arcade cascades in response to the primary wave, which reached the arcade at an oblique angle. Between 02:04 UT and 02:06 UT the wave front crosses a filament channel that is running parallel with the CH boundary. This filament is connected to the prominence that is seen over the south east (SE) limb in figure \ref{fig_cutout} (top left). After 02:06 the wave front subsequently decelerates to a velocity range of a few hundred km s$^{-1}$, which is consistent with the velocity measured in the FOV of EUVI-B. This front is clearly seen in figure \ref{fig_cutout} (top right) and the red mark is the position of the front as measured in the stack plot shown in figure \ref{fig_sliceAzoom}. There appears to be another secondary wave that emanates from the CH boundary occurring $\sim10$ minutes after the primary wave has past. It can be seen in the AIA 193 {\AA} running ratio movie between 02:10 UT and 02:30 UT propagating in the north east (NE) direction from the loop arcade (arrows in figure \ref{fig_cutout}). Its appearance seems to outline a loop system overlying a filament channel and oscillates with a period of $\sim 10$ minutes just like the loop arcade.

We note an effect that can arise in stack plots that are aligned between the active region that launched the coronal wave and the reflecting CH surface, such as slice ``B''. Looking closely at the AIA 211 {\AA} running ratio stack plot in figure \ref{fig_sliceB} (top) in the region between $-15^\circ$ and $-45^\circ$ there appears a feature after $\sim$02:05 UT which might be interpreted as a reflection. It has been highlighted in the figure as a dashed (green) line. this feature is in fact related to the oscillating loops at the CH boundary. Other examples in the literature of a stack plot aligned between the active region that launched the coronal wave and the reflecting CH surface are those presented by \citet{Gopalswamy_09} and \citet{LiT_12}. Upon close inspection of that region \citet{Attrill_10} explains that the reported reflection can be attributed to hot plasma channeled along loops in the direction back toward the active region. This scenario resembles the observation of the loops above the filament channel that we described above. If we had made a stack plot over this region it would in fact resemble a reflection. Just to be clear, what we described above \citet{Attrill_10} describes as hot plasma channeled along loops, in either case we are referring to an intensity enhancement that traces out along loops. Finally, in the case of \citet{LiT_12} the reported reflection occurs very close to the limb and after close inspection of that region it is difficult to tell if this could be attributed to hot plasma, density enhancement, a transverse/kink mode wave traveling along the loops and/or if this is a true reflection.

\subsection{Kinematics of the ``Transmitted Wave'' Front}\label{sec_transmit}

\begin{figure}
    \centerline{\hspace*{0.0\textwidth}
   \includegraphics[width=0.6\textwidth]{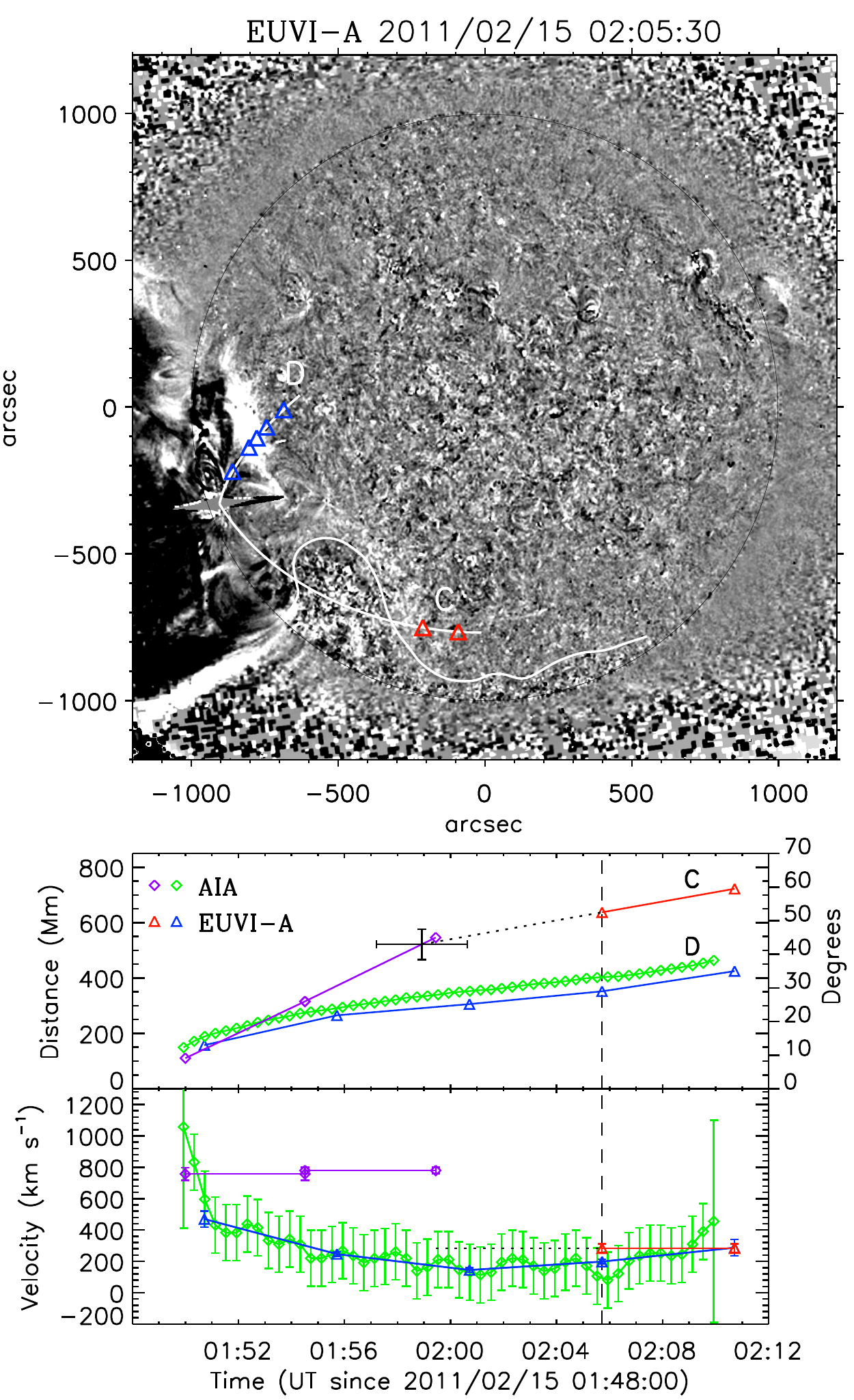}
   }
   \caption[]{Top panel is an EUVI-A 195 {\AA} running ratio frame taken at 02:05 UT. The coronal hole boundary outline is overlaid. The two great circle paths (C \& D) used to make the kinematic measurements are shown. The bottom two panels are the distance-time and velocity plots. Path C is the same as that plotted in figure \ref{fig_sliceC}, with AIA plotted piece-wise. Diamonds (green and violet) were measured with AIA observations, triangles (red and blue) were measured with EUVI-A.}
   \label{fig_transmit}
\end{figure}

Besides the reflection at the CH boundary, our full Sun movies show another interesting, and quite unique interaction between the wave and the coronal hole---a wave transmission. The top center and right panel of Figure \ref{fig_seq} show the appearance of a wave at the western part of the coronal hole, first seen at 02:05 UT within the FOV of EUVI-A (figure \ref{fig_transmit} top panel). This wave front wraps around the coronal hole boundary and lies clearly ahead of the EUV wave that propagates north west (NW) of the coronal hole. In other words, the part of the wave that goes through the coronal hole appears to \textit{accelerate\/} relative to the part that propagates in the quiet sun. This observation is fully consistent with the behavior of a fast-mode wave since the fast-mode speed will be higher in a coronal hole (due to its reduced plasma density) than in the quiet sun at a given height.

\begin{figure}
    \centerline{\hspace*{0.0\textwidth}
   \includegraphics[width=1\textwidth]{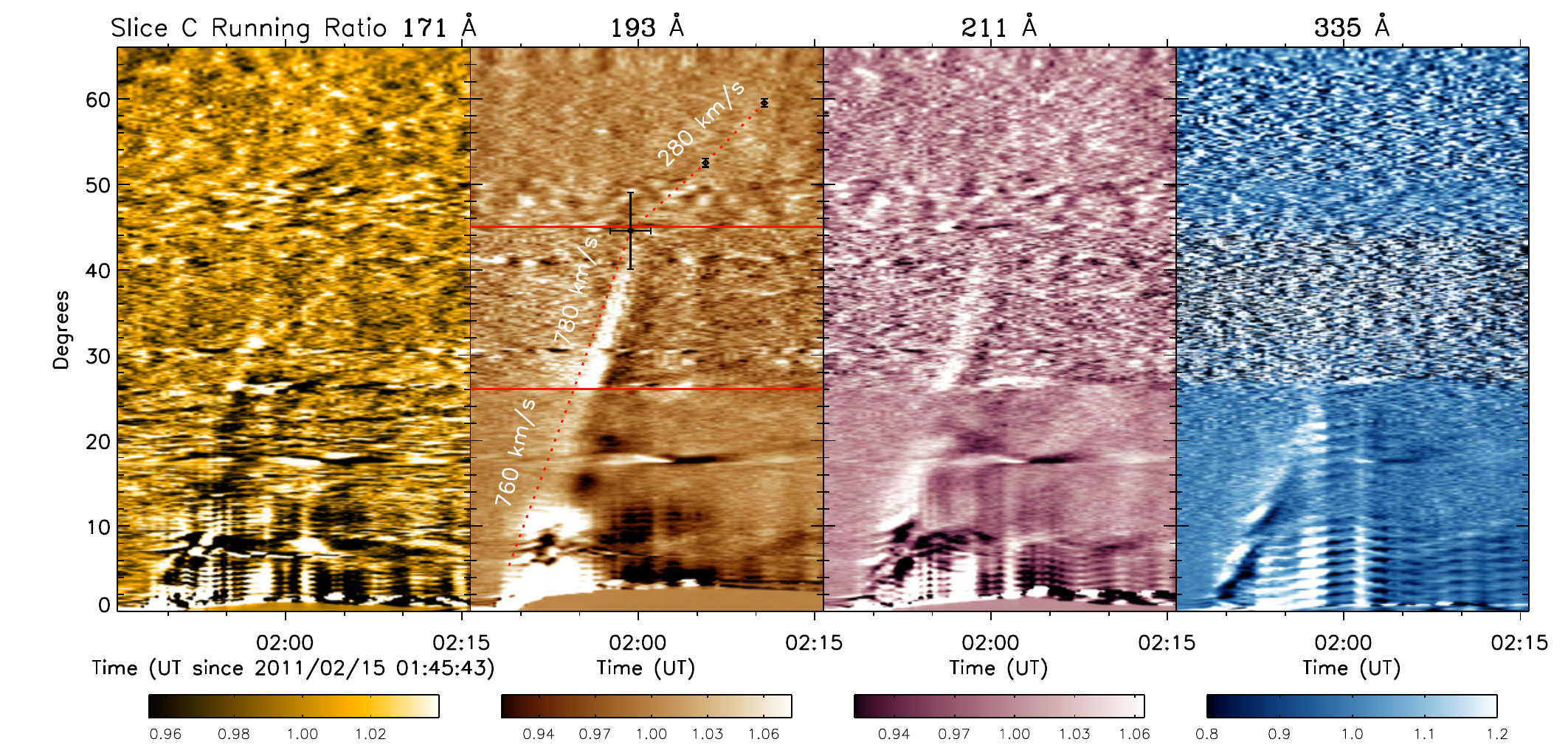}
   }
   \caption[]{Slice ``C'', that demonstrates transmission, originates at the flare site and passes over the coronal hole towards the south west. All four AIA channels are shown. The two red horizontal lines overlaid on the AIA 193 {\AA} plot at $26^\circ$ and $45^\circ$ indicated the bounds of the coronal hole.}
   \label{fig_sliceC}
\end{figure}

As was first introduced in section \ref{sec_primary} an apparent transmission of the primary wave is seen to occur across the CH. Here we take another great circle slice towards the south west (SW) originating at the flare site with a width of 18.6 arcsec within the FOV of AIA, labeled as ``C'' in figure \ref{fig_aia}. Stack plots of the four AIA channels are plotted together in figure \ref{fig_sliceC}. This slice passes over the CH region, bound by two horizontal red lines at $26^\circ$ and $45^\circ$ in the 193 {\AA} plot. The velocity of the front before reaching the CH boundary and through the CH are measured independently using the AIA 193 {\AA} observations. The velocity of the primary front was found to be $760 \pm 40$ km s$^{-1}$, and the front through the CH was $780 \pm 20$ km s$^{-1}$. As before, the error was estimated from making the fit ten times. These measurements do not seem to be significantly different and so it seems that the front propagates continuously across the CH boundary. The fact that the primary front velocity is comparable to the velocity within the CH can be understood as the primary front being initially driven by the CME to the CH  boundary, upon which time a freely propagating fast-mode wave traverses the CH, and that coincidentally the velocity of the primary front and the fast-mode speed within the CH are similar. This scenario fits within the context of the hybrid interpretation of EUV waves. It could be that the primary front observed is the CMEs outer envelope or pseudo-wave, as proposed by \citet{Schrijver_11}, and that upon reaching the CH a freely propagating fast-mode wave is launched through the CH, since pseudo-waves are not thought to cross into CHs.

The same scaling was applied throughout the AIA 193 {\AA} stack plot and it is seen that once the front reaches the CH its intensity increases by a few percent, probably because it is projected against the dimmer CH. This interpretation is further reinforced by the AIA 171 {\AA} observations which show a dimming outside the CH and a transition to an enhancement. The intensity enhancement is much smaller in 171 {\AA} because of the much small contrast in of CH in that AIA passband.  Because we are looking at the CH from an oblique angle towards the southern limb we are likely seeing the temporary pileup at the wave front against the faint CH. With the EUVI-A observations only two usable data points were found at 02:05 UT (52.5$^\circ$) and 02:10 UT (59.5$^\circ$), they are over plotted in figure \ref{fig_transmit} in which the front is clearly discernible SW of the CH. An error of $0.5^\circ$ was estimated for the two data points. A linear fit to these two data points yields a velocity $280 \pm 30$  km s$^{-1}$. The fit is extrapolated to meet the fit of the front through the CH as measured with AIA. Taking into account the error estimates, the EUVI-A fit intersect the AIA fit at $45^\circ \pm 5^\circ$ at 01:59 UT $\pm 1.7$ minutes. This point, with error bars, is plotted in figure \ref{fig_sliceC} along with the fits and the two EUVI-A data points. From this we conclude that what we are seeing within the CH is most likely the wave front and that upon exiting the CH and entering quiet Sun region it attains the velocity of the local fast-mode wave speed there.

Next, it is explicitly shown that the wave front towards the SW propagates ahead of the front toward the NW by comparing their kinematics. Slice C and a track towards the NW lying over quiet Sun region, labeled as ``D'' in figures \ref{fig_aia} and \ref{fig_transmit}, are compared. This track is assumed as a great circle originating at the flare site. The bottom two panels of \ref{fig_transmit} show the kinematics of both the C and D tracks. The track C measurements were described above and also plotted in figure \ref{fig_sliceC}, the (violet) diamonds are AIA piece-wise measurements, and the (red) triangles are the two data points measured with EUVI-A. Also plotted is the point where the EUVI-A fit intersects the AIA fit and the dotted line is the extrapolation. Along track D we were fortunate to have been able to make measurements in both AIA and EUVI-A. The (greed) diamonds are AIA, and the (blue) triangles are EUVI-A measurements. An error of 0.5$^\circ$ was estimated for both measurements. Looking at the plots it is seen that the two measurements from perpendicular vantage points are in agreement, especially in velocity (derived using 3-point Lagrangian interpolation). The discrepancy seen in the distance measurements can simply be attributed to projection effects between the two spacecrafts. Comparing the position of the wave front at 02:05:43 UT the front along the direction of track C has clearly propagated further away from the flare site than the front along track D by $\sim 285$ Mm or $\sim 23.5^\circ$. Furthermore, it is seen that tracks C and D have decelerated to comparable velocities on the order of $\sim300$ km s$^{-1}$.

\section{CONCLUSION \& DISCUSSION}\label{sec_conclusion}

We have demonstrated the wave nature of the coronal wave observed during the 2011 February 15 event by explicitly identifying and analyzing observations of secondary waves, which can be interpreted as reflection from and/or transmission through a CH. The asymptotic velocities, interpreted as reaching the local fast-mode speed, for all the wave ground track slices presented are comparable with each other (see table \ref{table_velo}). The difference in the asymptotic speeds of the ``reflected'' waves, slice ``A'', (380 vs. 280 km s$^{-1}$), of the meridional track toward the north, slice ``B'', (260 km s$^{-1}$) and of the SW track (280 km s$^{-1}$), that crossed the CH, seem to reflect the variations of the local fast-mode speed in the corona and are in agreement with global MHD models \citep[e.g.][]{Schmidt_10,Zhao_11}.  Furthermore, these differences may provide good diagnostics for coronal seismology \citep[e.g.][]{Yang_10,West_11}, but this topic is outside of the scope of this paper. Our results are also consistent with other authors, for example \citet{Warmuth_11} showed, based on a statistical study, that coronal waves with initial fast velocities ($v \ge 320$ km s$^{-1}$) show the greatest deceleration and attain final velocities between 200-300 km s$^{-1}$, \citet{Veronig_10} found an asymptotic velocity of a coronal wave far from its origin of $\sim280$ km s$^{-1}$, and \citet{Long_08,Long_11} found deceleration in the events that they studied. This can be explained if the coronal wave were in fact a freely-propagating fast-mode MHD wave \citep[e.g.][]{Warmuth_04,Veronig_10,Long_11,Warmuth_11}. \citet{Warmuth_11} have proposed that for fast events ($v \ge 320$ km s$^{-1}$) the physical nature of the coronal wave can be explained as being initially a large-amplitude nonlinear wave and/or shock, presumably driven by the CME, that subsequently evolves (and decelerates) to a linear fast-mode wave propagating at the characteristic wave speed. This interpretation is agreement with the hybrid view of coronal waves, in which the pseudo-wave can explain the outer envelope of the CME that drives a fast-mode wave that is left freely propagating close to the solar surface once the CME has propagated radially away. It is this fast-mode wave that we are referring to in this paper.

We have reported on features which appear to obey the law of reflection. Close inspection of the reflection to the east into the FOV of EUVI-B shows that the secondary wave observed may have been triggered by a cascading loop arcade. To the west, into the FOV of EUVI-A, a secondary wave propagates along a path that seems to obey the law of reflection. But, is what is seen to the west in fact an observation of reflection? Or a secondary wave launched by the active region or the CH resonating? It is difficult to tell with these observations because that region there are bright active region loops and the cadence of EUVI-A is not sufficient to resolve fine temporal structure in that region. Our work presented a case of reflection in which a different approach was taken by evoking the law of reflection and ray tracing possible trajectories of the wave towards and reflecting from the CH boundary. We used both running and base ratio stack plots as appropriate. While the reflection is a strong argument for the wave nature of coronal waves, it has been challenged as artifact of the data analysis method (\citet{Attrill_10} but see counter arguments in \citet{Zhukov_11}). When it comes to these techniques the primary problem arrises when analyzing a stack plot that is placed between the active region that launched the coronal wave and reflecting CH boundary surface. Two such possible effects may cause misleading effects to be seen in the stack plots. The first could be hot plasma, density enhancement, and/or a transverse/kink mode wave traveling along the loops aligned between the active region that launched the coronal wave and the CH boundary \citet{Attrill_10,Attrill_09}. The second could be loops triggered to oscillate along the CH boundary resulting in a signature in the stack plot that appears like reflection. Of course we must not be to hasty and dismiss the possibility that the wave did in fact reflect, the point we are trying to make is that each case of reflection should be considered carefully.

The transmitted wave toward the SW can be identified in both running and base ratio images without ambiguity because it propagates across ``pristine'' quiet Sun. This is the first time, to our knowledge, that such a feature is identified in observations. The observation of the coronal hole crossing (or transmission) is a very strong argument for the wave nature of these EUV disturbances. Until now, it has been reported based on empirical evidence that coronal waves do not traverse coronal holes \citep[e.g.][]{Thompson_99,Veronig_06,Attrill_07b,Ma_09}. However, transmission has been discussed in the simulations of \citet{Schmidt_10}, who also discuss it in terms of waves launched by the ``resonating'' of the CH. Nevertheless we think that the same effect (e.g. transmission) is occurring in this event (either by resonance is not clear nor could it be determined based on these observations alone). Furthermore, we do see secondary waves launched all along the boundary of that CH (see provided movies), even beyond the reach of the original coronal wave, meaning that those waves would have had to cross the CH. But why it hasn't been observed before? We think that the size of the coronal hole is likely the determinant factor. If the coronal hole is large then the resonance may die out before it reaches the side opposite to the wave impact. On the other hand, if a hole is small the resonance may be sufficient to launch the secondary wave on the opposite side or all around the hole as suggested by \citet{Schmidt_10} and seen in this event as well as (partially) seen in the 2007 May 19 event \citep{Gopalswamy_09}.

The reflection and transmission observations are hard to reconcile with a pure pseudo-wave interpretation and our results provide further evidence in favor for the wave nature of coronal waves. But note that it is the fast-mode wave that we are referring to originally driven by the CME, whose outer envelope may be described by the pseudo-wave. It may be the case that for weaker events a fast-mode wave is not triggered or is triggered but has an amplitude that does not produced enough heating to be detected, in these cases a pure pseudo-wave interpretation may be appropriate as the CME and its outer envelope may be the only signatures observed. In other similar events to the one studied here: the 2010 July 27 event reported by \citet{Chen_11}, the 2011 June 7 event reported by \citet{Cheng_12}, and the 2010 June 13 event reported by \citet{Downs_12}, it is shown that coronal wave events are a composite phenomenon comprised of a CME that drives a fast-mode wave, giving evidence to the hybrid interpretation. Based on those results what we have presented here is a detailed account of the wave component, initially driven by the CME, and that a pseudo-wave component should exist but the point at which one can distinguish between the two is not readily possible only to say that it should be early on in order to account for the secondary wave effects presented. We estimate that for this event the separation should have occurred some time before 02:00 UT based on the fact that the transmission is observed to occur after that time, and transmission is purely a wave effect.

A projection effect of the CME material over the CH is the most obvious suggestion for the EUV emission inside the CH. Which could be the projection of deflected coronal material by the CME towards the south. But the CME material propagates away from the CH (towards the ecliptic plane) and its southern flank stops at the boundary \citep{Schrijver_11}. Even if we assume that the expanding material is somehow tilted to the SW (as seen from Earth), it is difficult to account for the appearance of a propagating low coronal feature outside of the CH to the SW (seen in the FOV of EUVI-A). The CME field lines cannot penetrate through the CH. No such observations has even been reported in countless limb observations of CMEs. The reflection is an additional problem. There is no secondary CME, we use both running and base ratio images and we do not track the reflection back through the incident path (as was done by \citet{Gopalswamy_09} and \citet{LiT_12}), so there is no chance for confusion. Secondary waves can also be seen in AIA observations of the CME and coronal wave event on 2011 June 7 \citep{LiT_12} and a transmitted wave can be seen in the 2011 February 24 CME.

Instead, the fast-mode interpretation neatly accounts for all observations. Wave reflection is expected due to the sharp change in plasma parameters at the coronal hole/quiet sun interface as has been simulated \citep[][]{Schmidt_10} and reported observationally \citep{Long_08,Gopalswamy_09,LiT_12}, and wave transmission is predicted by MHD simulations \citep[][]{Schmidt_10}. Moreover oscillations of loops have also been reported as due to primary coronal wave fronts \citep[e.g.][]{Aschwanden_11}, which, as was shown here, can trigger secondary waves interpreted as being fast-mode MHD. 

In summary, we have undertaken a detailed analysis of the global kinematics of an EUV wave using the full Sun coverage afforded by the STEREO and SDO missions and have found several important results:
\begin{itemize}
\item Secondary wave effects attributed to the interaction of the primary coronal wave with the CH were described, including reflection and the first detection of wave transmission though a CH.
\item We report detailed observations of the interaction of a coronal wave with a loop arcade that straddles the boundary of a CH, which subsequently launches a secondary wave, whose trajectory resembles the law of reflection. Such observations have not been reported before and suggest an origin to secondary waves that appear as reflections launched by CHs.
\item Our observations and kinematic analysis is fully consistent with MHD simulations of fast-mode waves and suggests that these waves are initially driven, presumably by the expanding CME, but eventually relax to a freely propagating wave traveling at the local fast-mode speed. 
\item Our analysis shows that the quite Sun MHD wave speed can have modest variations. 
\item We also report the first global measurements of the angular extent of EUV waves, which reached at least or approximately $200^\circ$ before becoming too diffuse to be detected.
\end{itemize}

\acknowledgements

We thank the anonymous referee for their constructive comments that have greatly improved this work. We thank Guillermo Stenborg for providing the AIA wavelet enhanced images. O.O. acknowledges valuable discussions with Cooper Downs. This research was performed while O.O. held a National Research Council Research Associateship Award at Naval Research Laboratory. The work of O.O. and A.V. is funded by NASA contract S-136361-Y. J.Z. is supported by NSF grant ATM-0748003 and NASA grant NNX07AO72G. X.C. is supported by NSFC under grants 10673004, 10828306, and 10933003 and NKBRSF under grant 2011CB811402. The AIA data used here are courtesy of SDO (NASA) and the AIA consortium. We thank the AIA team for the prompt and easy access to calibrated data. The SECCHI data are produced by an international consortium of the NRL, LMSAL and NASA GSFC (USA), RAL and Univ. Bham (UK), MPS (Germany), CSL (Belgium), IOTA and IAS (France).

\bibliographystyle{plainnat}

%%%%%%%% Table 1 %%%%%%

\begin{table}
    \begin{tabular}{|l|l|l|}
        \hline
        ~       & Initial                                                                      & Asymptotic                 \\ \hline
        Slice A & 1500 km s$^{-1}$ East (See Figures \ref{fig_sliceAzoom}  \&  \ref{fig_trigger}) & 280 km s$^{-1}$ East       \\ 
        ~       & ~                                                                            & 380 km s$^{-1}$ West       \\  \hline
        Slice B & 720 km s$^{-1}$ North                                                        & 260 km s$^{-1}$ North      \\ 
        ~       & 740 km s$^{-1}$ South                                                        & ~                          \\ \hline
        Slice C & 760 km s$^{-1}$ South West                                                   & 280 km s$^{-1}$ South West \\ 
        ~       & 780 km s$^{-1}$ Transmitted                                                  & ~                          \\
        \hline
    \end{tabular}
    \caption[Table of Velocities]{Measured Velocities along the three ground track slices. Initial refers to the initially velocity, for Slice A this refers to the velocity of the wave upon being launched by the cascading loops, and Slice B and Slice C refer to the velocity of the primary wave soon after the flare onset. Asymptotic refers to the velocity attained after some time up to the point where the wave is no longer measurable.}
    \label{table_velo}
\end{table}

%%%%%%%%% figure 1 %%%%%%

%%%%%%%%% figure 2 %%%%%%

%%%%%%%%% figure 3 %%%%%%

%%%%%%%%% figure 4 %%%%%%

%%%%%%%%% figure 5 %%%%%%

%%%%%%%%% figure 6 %%%%%%

%%%%%%%%% figure 7 %%%%%%

%%%%%%%%% figure 8 %%%%%%

%%%%%%%%% figure 9 %%%%%%

%%%%%%%%% figure 10 %%%%%%

%%%%%%%%% figure 11 %%%%%%

%%%%%%%%% figure 12 %%%%%%

\end{document}